\documentclass[journal]{IEEEtran}
\ifCLASSINFOpdf
\else
   \usepackage{graphicx}
   \graphicspath{{../eps/}}
   \DeclareGraphicsExtensions{.eps}
\fi

\usepackage{amsmath}
\usepackage{amsfonts,amssymb}
\usepackage{amssymb}
\usepackage{wrapfig}
\usepackage{psfrag}
\usepackage{epstopdf}
\usepackage{cite}
\usepackage{graphicx}
\usepackage{subfigure}
\usepackage{threeparttable}
\usepackage{cases}
\usepackage{subeqnarray}
\usepackage{color}
\usepackage{underscore}
\usepackage{verbatim}
\usepackage{bm}
\usepackage{stfloats}
\usepackage{xpatch}
\usepackage{makecell}
\newtheorem{theorem}{Theorem}
\newtheorem{lemma}{Lemma}

\newtheorem{algorithm}{Algorithm}
\newtheorem{proposition}{Proposition}

\usepackage{algorithm}
\usepackage{algorithmic}
\usepackage[table]{xcolor}

\begin{document}
\title{Wireless Communication for Low-Altitude Economy with UAV Swarm Enabled\\ Two-Level Movable Antenna System}
%
%
%
\author{Haiquan~Lu,~\IEEEmembership{Member,~IEEE,}
        Yong~Zeng,~\IEEEmembership{Fellow,~IEEE,}
        Shaodan~Ma,~\IEEEmembership{Senior Member,~IEEE,}\\
        Bin~Li,~\IEEEmembership{Senior Member,~IEEE,}
        Shi~Jin,~\IEEEmembership{Fellow,~IEEE,}
        and
        Rui~Zhang,~\IEEEmembership{Fellow,~IEEE}
\thanks{The work of Yong Zeng was supported in part by the National Natural Science Foundation of China under Grant 62571116, and in part by the Mobile Information Networks-National Science and Technology Major Project under Grant 2025ZD1304500. The work of Shaodan Ma was supported in part by the Science and Technology Development Fund, Macau SAR under Grants 0114/2025/AMJ, 0020/2025/RIB1, 001/2024/SKL, and 0002/2025/EQP, and in part by the Research Committee of University of Macau under Grant MYRG-GRG2025-00143-IOTSC. The work of Rui Zhang was supported in part by the National University of Singapore under Grant A-8003646-00-00 and Grant A-8003676-00-00, in part by the National Natural Science Foundation of China under Grant 62331022, and in part by Guangdong Major Project of Basic and Applied Basic Research under Grant 2023B0303000001. Part of this work has been presented at the 2025 IEEE ICCC Workshop, Shanghai, China, 10-13 August 2025 ~\cite{lu2025enabling}. (\emph{Corresponding author: Yong Zeng.}) }
\thanks{Haiquan Lu is with the School of Electronic and Optical Engineering, Nanjing University of Science and Technology, Nanjing 210094, China (e-mail: haiquanlu@njust.edu.cn).}
\thanks{Yong Zeng is with the National Mobile Communications Research Laboratory, Southeast University, Nanjing 210096, China, and is also with the Purple Mountain Laboratories, Nanjing 211111, China (e-mail: yong_zeng@seu.edu.cn). }
\thanks{Shi Jin is with the School of information Science and Engineering, Southeast University, Nanjing 210096, China (e-mail: jinshi@seu.edu.cn).} 
\thanks{Shaodan Ma is with the State Key Laboratory of Internet of Things for Smart City and the Department of Electrical and Computer Engineering, University of Macau, Macao SAR, China (e-mail: shaodanma@um.edu.mo).}
\thanks{Bin Li is with School of Aeronautics and Astronautics, Sichuan University, Chengdu, Sichuan 610065, China (e-mail: bin.li@scu.edu.cn).}
\thanks{Rui Zhang is with School of Science and Engineering, Shenzhen Research Institute of Big Data, the Chinese University of Hong Kong, Shenzhen, Guangdong 518172, China (e-mail: rzhang@cuhk.edu.cn). 
}
%
}

\maketitle

 \begin{abstract}
 Unmanned aerial vehicle (UAV) is regarded as a key enabling platform for low-altitude economy, due to its advantages such as three-dimensional (3D) maneuverability, flexible deployment, and line-of-sight (LoS) air-to-air/ground communication links. In particular, the intrinsic high mobility renders UAV especially suitable for operating as a movable antenna (MA) from the sky. In this paper, by exploiting the flexible mobility of UAV swarm and antenna position adjustment of MA, we propose a novel UAV swarm enabled two-level MA system, where UAVs not only individually deploy a local MA array, but also form a larger-scale MA system with their individual MA arrays via swarm coordination. We formulate a general optimization problem to maximize the minimum achievable rate over all ground user equipments (UEs), by jointly optimizing the 3D UAV swarm placement positions, their individual MAs' positions (or local positions), and receive beamforming for different UEs. To gain useful insights, we first consider the special case where each UAV has only one antenna, under different scenarios of one single UE, two UEs, and arbitrary number of UEs. In particular, for the two-UE case, we derive the optimal UAV swarm placement positions in closed-form that achieves inter-UE interference (IUI)-free communication when the uniform plane wave (UPW) model holds, where the UAV swarm forms a uniform sparse array (USA) satisfying minimum safe distance constraint. While for the general case with arbitrary number of UEs, we propose an efficient alternating optimization algorithm to solve the formulated non-convex optimization problem. Then, we extend the results to the case where each UAV is equipped with multiple antennas. Numerical results verify that the proposed low-altitude UAV swarm enabled MA system significantly outperforms various benchmark schemes, thanks to the exploitation of two-level mobility to create more favorable channel conditions for multi-UE communications.
 \end{abstract}


\IEEEpeerreviewmaketitle
\section{Introduction}
 Low-altitude economy has emerged as a new integrated economic form, involving assorted low-altitude activities within sub-1,000-meter airspace domains \cite{jiang20236g,cheng2025networked}. By leveraging manned/unmanned aerial vehicles (UAVs), and electric vertical take-off and landing (eVTOL) aircrafts, low-altitude economic ecosystems begin to flourish, which encompass a wide range of applications like logistics and delivery, agricultural plant protection, cultural tourism, environmental monitoring, and so on. In particular, benefiting from the advantages such as three-dimensional (3D) maneuverability, flexible deployment, and line-of-sight (LoS) air-to-air/ground communication links, UAV is regarded as an indispensable part of the low-altitude economy, which has driven various applications in wireless communication and sensing \cite{zeng2019accessing,mozaffari2019tutorial,song2024overview}. Specifically, from the perspective of wireless communication, there exist two paradigms, i.e., UAV-assisted wireless communication and cellular-connected UAV communication \cite{zeng2019cellular}. For example, UAV equipped with a base station (BS) or relay can be deployed on demand to assist in the conventional cellular wireless networks, so as to satisfy the ubiquitous connectivity required by the future sixth-generation (6G) wireless network. Extensive research efforts have been devoted to this direction, including UAV channel and energy consumption modeling, performance analysis, and placement/trajectory design \cite{zeng2019accessing,mozaffari2019tutorial}. On the other hand, the role of UAV can be shifted from the aerial BS/relay to aerial user equipments (UEs), thus enabling a new paradigm where both the ground and aerial UEs coexist in future wireless networks \cite{zeng2019cellular,zhang2019cellular,zeng2019accessing}. Moreover, as integrated sensing and communication (ISAC) has been identified as one of the six major usage scenarios for 6G \cite{ITU}, research on ISAC with UAV has gained a surge of interest recently \cite{mu2023uav,pan2024cooperative,jing2024isac}. Similarly, UAV can act as an aerial anchor for providing the ISAC service from the sky, or an aerial target to be sensed \cite{song2024overview}. 
 
 However, UAVs are subject to various practical constraints in terms of size, weight, dynamics, and power, rendering a single UAV challenging to execute sophisticated communication or sensing missions \cite{zeng2019accessing,javed2024state,li20223d,li2024joint,zhang2025joint}. To tackle this issue, UAV swarm, i.e., a group of coordinated UAVs, can cooperatively accomplish sophisticated tasks, e.g., large-scale data collection and environment monitoring. Compared to single UAV, UAV swarm is able to significantly improve the overall payload capability and expand the communication and sensing coverage area. Moreover, the cooperative mechanism helps to reduce the task completion time and offers an enhanced robustness against anomalies or failures than single UAV, since the task of any malfunctioning UAV can be transferred to its neighboring functioning UAVs \cite{javaid2023communication}. Given the above promising advantages, extensive research endeavours have been devoted to UAV swarm assisted communication and sensing. For example, in \cite{fan2019channel}, an efficient channel estimation and self-positioning approach was proposed for UAV swarm communication with unknown displacements among UAVs. The authors in \cite{mou2021deep} proposed a 3D irregular terrain area coverage scheme with UAV swarm, and the results show that the proposed scheme can cover the area with little redundancy. Besides UAV swarm assisted communication, recent research works have studied the integration of ISAC with UAV swarm, since the LoS-dominating channel is particularly favorable for sensing \cite{song2024overview}. Research on UAV swarm ISAC involves various performance metrics, including signal-to-interference-plus-noise ratio (SINR) and achievable rate for communication \cite{cheng2025networked,jing2024isac}, radar mutual information \cite{liu2024fair} and Cram\'{e}r-Rao lower bound (CRLB) for sensing \cite{pan2024cooperative,jing2024isac,wang2024cooperative}. Despite the promising advantages, UAV swarm also involves several issues, such as the latency caused by information exchange among UAVs, complex collision avoidance planning, and high-precision localization and sensing for UAV swarm regulation \cite{xu2025integrated,javed2024state}. In particular, to tackle the challenge of accurately localizing and sensing the densely located swarm UAVs, a super-resolution communication and sensing method was proposed for UAV swarm in \cite{xu2025integrated}.

 It is worth mentioning that the inherent flexible mobility of UAV renders it particularly suitable for operating as a movable antenna (MA) \cite{zhu2024movable,zhu2023full,ma2024mimo,zhu2025tutorial,lu2024group,dong2024movable,shao20256dModeling,shao20256ddiscrete,shao2025distributed,shao2025tutorial} or fluid antenna system (FAS) \cite{wong2021fluid,new2024tutorial} from the sky. For example, each UAV may be equipped with a single or multiple antennas, and a flexible antenna array architecture can be enabled via adjusting the UAV swarm topology. In contrast to conventional fixed-position antenna (FPA) where inter-antenna spacing is usually fixed as half wavelength, MA dynamically adjusts the antenna position via mechanical and/or electrical control, so as to pursue a favorable channel condition and circumvent the deep fading scenario \cite{zhu2024movable,wong2021fluid}. This thus achieves performance improvement with optimized antenna positions, which may be a complement to extremely large-scale multiple-input multiple-output (MIMO) technique whereby the spectral efficiency and spatial resolution are significantly enhanced via the orders-of-magnitude scaling of antennas \cite{lu2022near,lu2024tutorial}. Subsequently, antenna mobility is extended to a more general concept, where both the antenna position and rotation can be flexibly adjusted, i.e., six-dimensional movable antenna (6DMA) \cite{shao20256dModeling,shao20256ddiscrete,shao2025distributed,shao2025tutorial}. Driven by such new design degrees of freedom (DoFs), a large body of literature focuses on this technology and demonstrates its performance gain in communication outage probability, spectral efficiency and network capacity, as well as sensing accuracy (see \cite{zhu2025tutorial,new2024tutorial,shao2025tutorial} and references therein). Besides active MA, the authors in \cite{lu2025wireless} further proposed a new flexible passive reflector architecture, enabling the flexible beamforming direction adjustment via reflector placement and rotation angle optimization.

 Note that compared to MAs, UAVs possess even more flexible and wider-range movement, and thus an interesting new  idea is to utilize UAV swarm to form MA array. Moreover, different from existing research on MA that only optimizes antenna positions, UAV swarm enabled MA can fully unlock its mobility advantage, where the array geometry is able to be dynamically reconfigured via trajectory optimization, thus yielding a varying array geometry. Preliminary efforts have been devoted to the single UAV-mounted MA \cite{bai2024movable,tang2024uav,liu2025uav,ren20256d}. Specifically, a directional MA was mounted on a UAV to minimize the total data collection time in a backscatter sensor network in \cite{bai2024movable}. In \cite{tang2024uav} and \cite{liu2025uav}, the authors utilized the single UAV-mounted MA array to facilitate the minimum beamforming gain for secondary users (SUs) and enhance the sum rate of all UEs, respectively. In \cite{ren20256d}, the 6DMA was deployed at the single UAV to mitigate the interference caused by the co-channel terrestrial transmissions. Besides the active antenna, the authors in \cite{liu2024uav} utilized the single UAV-mounted passive 6DMA, i.e., by carrying an intelligent reflecting surface (IRS) on a UAV \cite{lu2021aerial}, to maximize the minimum received signal-to-noise ratio (SNR) among all UEs. Nevertheless, these works neither take into account the UAV swarm scenario, nor explore the capability of MA system enabled by UAV swarm. Meanwhile, it is also worth pointing out that the UAV swarm enabled MA system may face several practical challenges, such as the limited endurance of UAV swarm due to the finite onboard battery, positional inaccuracy, array misalignment, orientation jitter, and synchronization issues, which deserve further investigations.
  
 In this paper, motivated by the inherent mobility of UAV swarm and antenna position adjustment of MA, we propose a novel UAV swarm enabled two-level MA system. Specifically, by individually deploying a local MA array on each UAV, all the spatially distributed UAVs cooperatively construct a larger-scale MA system according to their topology. Compared to existing MA systems, UAV swarm enabled MA system involves the dual-scale antenna spatial movement, i.e., the small-scale antenna movement within each local MA array, e.g., at wavelength level, and the large-scale UAV movement, so as to better harness the spatial variations of wireless channels in different scales. In particular, the UAV swarm enabled MA system can be deployed on demand, which is appealing for assorted application scenarios, such as information dissemination and data collection, data offloading in hotspot areas, and rapid communication restoration after infrastructure failure \cite{zeng2019accessing}. Moreover, benefiting from the desired LoS air-to-air/ground links, UAV swarm enabled MA system can provide the ISAC service efficiently, so as to support low-altitude economy applications such as logistics and delivery as well as environmental monitoring. In this paper, by considering low-altitude UAV swarm enabled uplink multi-UE communication, we aim to maximize the minimum achievable rate over all ground  UEs. The main contributions of this paper are summarized as follows.

 \begin{itemize}[\IEEEsetlabelwidth{12)}]
 \item First, we propose a novel low-altitude UAV swarm enabled two-level MA system, where each UE channel depends on both UAV swarm placement positions and their mounted MA array positions. Subject to the minimum safe distance constraint for UAVs \cite{zeng2019accessing}, as well as mutual coupling avoidance constraint for UAV-mounted MA array, we formulate an optimization problem to maximize the minimum achievable rate over all ground UEs, by jointly optimizing the 3D UAV swarm placement positions, their individual MAs' positions, and receive beamforming for all UEs. 
 \item Second, to gain useful insights, we consider the special case of single-antenna UAV. It is shown that for the single UE communication, the resulting SNR after applying the optimal maximal-ratio combining (MRC) beamforming only depends on the number of UAVs, irrespective of their array geometry. Then, for the two-UE communication, we derive the optimal UAV swarm placement positions that achieves the maximum communication rate without suffering any inter-UE interference (IUI) when the uniform plane wave (UPW) model holds. The result shows that under the minimum safe distance constraint, UAV swarm forms a uniform sparse array (USA) \cite{li2025sparse} for the IUI-free communication. Moreover, for arbitrary number of UEs, we propose an efficient alternating optimization algorithm to solve the highly non-convex problem, where the optimal receive beamforming is derived in closed-form for given UAV swarm placement positions, and for given receive beamforming, the UAV swarm placement optimization problem can be solved via the successive convex approximation (SCA) technique \cite{zeng2019accessing}.
 \item Last, we consider the general case of multi-antenna UAV, where each antenna can move independently within the specified region at its mounted UAV. For the single and two-UE communications, similar results can be obtained as the scenario of single-antenna UAV. In particular, we also consider the special case where all the UAVs adopt the USA \cite{wang2024enhancing}, the IUI-free communication can be achieved in the two-UE case via joint UAV placement positions and sparsity level adjustment. Next, the alternating optimization method is applied to solve the problem with arbitrary number of UEs. Subsequently, we extend the analysis to the practical case by considering synchronization and position errors, where a channel estimation-based error compensation scheme is proposed. Numerical results are presented to validate the superior performance of the proposed UAV swarm enabled MA system over various benchmarks, thanks to the two-level mobility to create more favorable channel conditions. 
 \end{itemize}

 The remainder of the paper is organized as follows. Section~\ref{sectionSystemModel} presents the system model of a low-altitude UAV swarm enabled MA system and formulates the problem to maximize the minimum achievable rate among UEs. Section~\ref{sectionSingleAntennaUAV} considers the special case of single-antenna UAV, and the result is extended to the general case of multi-antenna UAV in Section~\ref{sectionMultiAntennaUAV}. Section~\ref{sectionNumericalResult} presents the numerical results to validate the performance of UAV swarm enabled MA systems. Finally, we conclude the paper in Section~\ref{sectionConclusion}.

 \emph{Notations:} Scalars are denoted by italic letters. Vectors and matrices are denoted by bold-face lower- and upper-case letters, respectively. The spaces of $M \times N$ complex-valued matrices are denoted as ${{\mathbb{C}}^{M \times N}}$. For a vector ${\bf{x}}$, $\left\| {\bf{x}} \right\|$ denotes its Euclidean norm. For an arbitrary-size matrix ${\bf A}$, its complex conjugate, transpose, and Hermitian transpose are denoted by ${\bf A}^*$, ${\bf A}^T$, ${\bf A}^H$, respectively. The spectral norm and Frobenius norm of ${\bf A}$ are denoted as ${\left\| {\bf{A}} \right\|_2}$ and ${\left\| {\bf{A}} \right\|_F}$, respectively. The distribution of a circularly symmetric complex Gaussian random vector with mean $\bf{x}$ and covariance matrix $\bf{\Sigma}$ is denoted as ${\cal CN}\left( {\bf{x},\bf{{\Sigma}}} \right)$, and $\sim$ stands for ``distributed as". For a complex-valued number $x$, ${\mathop{\rm Re}\nolimits} \left\{ x \right\}$ represents its real part. $\bmod \left( {a,b} \right)$ represents the integer modulo operation, which returns the remainder after division of $a$ by $b$. The symbol ${\rm j}$ denotes the imaginary unit of complex numbers, with ${{\rm j}^2} =  - 1$. ${\cal O}\left({\cdot}\right)$ denotes the standard big-O notation.
\section{System Model And Problem Formulation}\label{sectionSystemModel}
 \begin{figure}[!t]
 \centering
 \centerline{\includegraphics[width=3.1in,height=2.6in]{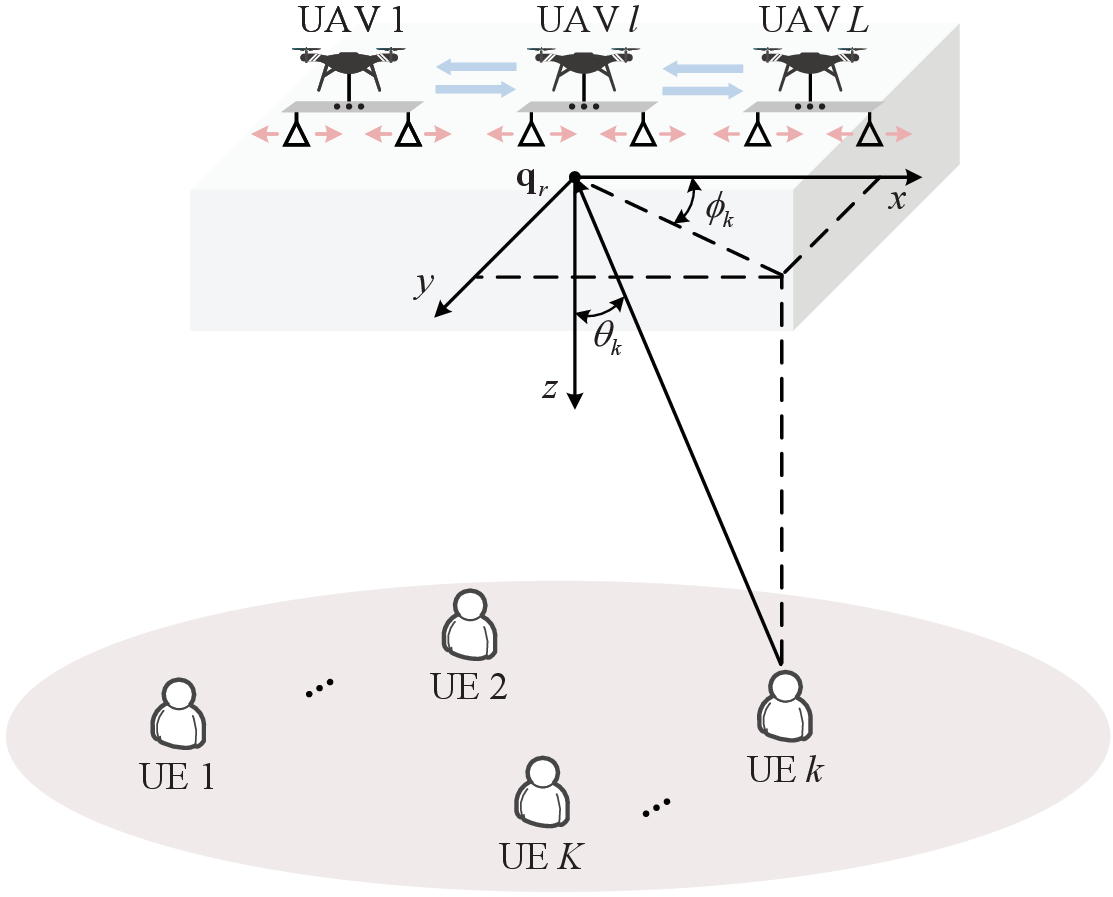}}
 \caption{Wireless communication with a low-altitude UAV swarm enabled two-level MA system.}
 \label{fig:systemModel}
 \end{figure}

 As illustrated in Fig.~\ref{fig:systemModel}, we consider a low-altitude UAV swarm enabled wireless communication system, where $L$ UAVs cooperatively serve $K$ single-antenna ground UEs. Each UAV is equipped with a linear MA array with $M$ antenna elements, while a flexible two-level MA system is formed by dynamically moving each UAV and the positions of its individual MAs. Without loss of generality, we consider a 3D Cartesian coordinate system, where the location of UE $k$ is denoted as ${\bf w}_k = \left[{{x_k}, {y_k}, 0  } \right]^T$, $k \in {\cal K}$, with ${\cal K} \triangleq \left[1, \cdots, K\right]$. Let the first antenna be the reference point for each UAV, with the coordinate of UAV $l$ denoted as ${{{\bf{\bar q}}}_l} = {\left[ {{{\bar x}_l},{{\bar y}_l},{{\bar z}_l}} \right]^T}$, $1 \le l \le L$. Moreover, the coordinate of antenna $m$ for UAV $l$ is ${{\bf{\bar q}}_{l,m}} = {{\bf{\bar q}}_l} + {{\bf{b }}_{l,m}}$, where ${{\bf{b }}_{l,m}} \triangleq  {\left[ { {{b _{l,m}}},0,0} \right]^T}$ and we assume ${b _{l,1}} = 0$, $\forall l$, without loss of generality. As such, the position of antenna $m$ can be dynamically changed by adjusting ${b _{l,m}}$ for UAV $l$. This thus enables a two-level MA system, realized by the joint control of UAVs' movement and their individual antennas' movement. 
 
 Let ${{\bf{q}}_r} = {\left[ {{x_r},{y_r},{z_r}} \right]^T}$ denote the reference location for the $L$ UAVs. Let ${{\bm{\kappa }}_k} = {\left[ {{\Phi _k},{\Psi _k},{\Theta _k}} \right]^T}$ denote the wave propagation direction from UE $k$ to ${\bf q}_r$, with ${\Phi _k} \triangleq \sin {\theta _k}\cos {\phi _k}$, ${\Psi _k} \triangleq \sin {\theta _k}\sin {\phi _k}$, and ${\Theta _k} \triangleq \cos {\theta _k}$, respectively, where ${{\theta _k}}$ and ${{\phi _k}}$ denote the elevation and azimuth angles of arrival (AoAs), respectively, as illustrated in Fig.~\ref{fig:systemModel}. Moreover, let ${{\bf{q}}_l} = {\left[ {{x_l},{y_l},{z_l}} \right]^T}$ denote the position of UAV $l$ relative to the reference point ${{\bf{q}}_r}$, and it follows that ${{{\bf{\bar q}}}_l} = {{\bf{q}}_r} + {{\bf{q}}_l}$. Similarly, we have ${{{\bf{\bar q}}}_{l,m}} = {{\bf{q}}_r} + {{\bf{q}}_{l,m}}$, with ${{\bf{q}}_{l,m}} = {{\bf{q}}_l} + {{\bf{b}}_{l,m}}$ denoting the position of antenna $m$ for UAV $l$ relative to the reference point ${{\bf{q}}_r}$. We assume that each UAV knows its relative position to the reference point, which can be obtained via satellite-based localization systems, e.g., global positioning system (GPS) or Beidou.
 
 In practice, the minimum safe distance constraints need to be satisfied by UAVs, which is given by \cite{zeng2019accessing}
 \begin{equation}\label{collisionAvoidanceConstraint}
 \left\| {{{\bf{q}}_l} - {{\bf{q}}_{l'}}} \right\| \ge {d_{\min }},\ \forall l, l' \ne l,
 \end{equation}
 where $d_{\min}$ denotes the minimum distance to guarantee the safe operation between UAVs. Besides, to avoid the mutual coupling between adjacent MA elements at each UAV, the minimum distance ${\tilde d}_{\min }$ is required, which yields the following constraints, 
 \begin{equation}\label{MCAvoidanceConstraint}
 \left\| {{{\bf{q}}_{l,m}} - {{\bf{q}}_{l,m'}}} \right\| = \left| {{b _{l,m}} - {b _{l,m'}}} \right| \ge {{\tilde d}_{\min }}, \forall l,m,m' \ne m,
 \end{equation}
 where ${\tilde d}_{\min }$ is in practice much smaller than $d_{\min}$.

 Note that the communication links between UAVs and ground UEs are of high probability to be over LoS channels in practice. We consider the general near-field spherical wavefront model for the channels between UAVs and ground UEs, which includes the far-field UPW model as a special case. Meanwhile, the signals from UE to the local MA array of each UAV can be well approximated as UPW, since the movable region of the local MA array is much smaller than the link distance. Let ${{\bm{\kappa }}_{k,l}} = {\left[ {{\Phi _{k,l}},{\Psi _{k,l}},{\Theta _{k,l}}} \right]^T}$ denote the wave propagation direction from UE $k$ to ${{\bf{\bar q}}_l}$, where ${\Phi _{k,l}} = \sin {\theta _{k,l}}\cos {\phi _{k,l}}$, ${\Psi _{k,l}} = \sin {\theta _{k,l}}\sin {\phi _{k,l}}$, and ${\Theta _{k,l}} = \cos {\theta _{k,l}}$, with ${\theta _{k,l}}$ and ${\phi _{k,l}}$ being the elevation and azimuth AoAs, respectively. For UAV $l$, the difference of the wavefront propagation distance between ${{\bf{\bar q}}_l}$ and ${{\bf{\bar q}}_{l,m}}$ for UE $k$ is 
 \begin{equation}
 {\varphi _k}\left( {{{{\bf{\bar q}}}_{l,m}}} \right) = {\bm{\kappa }}_{k,l}^T\left( {{{{\bf{\bar q}}}_{l,m}} - {{{\bf{\bar q}}}_l}} \right) = {\bm{\kappa }}_{k,l}^T{{\bf{b}}_{l,m}} = {\Phi _{k,l}}{b_{l,m}},
 \end{equation}
 which is the projection length of the vector $\left( {{{{\bf{\bar q}}}_{l,m}} - {{{\bf{\bar q}}}_l}} \right)$ onto the wave propagation direction vector ${{\bm{\kappa }}_{k,l}}$. Then, the receive array response vector of UAV $l$ for UE $k$ is given by \cite{lu2024tutorial}
 \begin{equation}\label{arrayResponseVectorSingleUAV}
 \begin{aligned}
 & {{\bf{a}}_{k,l}}\left( {\left\{ {{{{\bf{\bar q}}}_{l,m}},\forall m} \right\}} \right)= {e^{ - {\rm{j}}\frac{{2\pi }}{\lambda }{r_{k,l}}}} \times \\
 &\ \ \ \ \ \ {\left[ {{e^{ - {\rm{j}}\frac{{2\pi }}{\lambda }{\varphi _k}\left( {{{{\bf{\bar q}}}_{l,1}}} \right)}}, \cdots ,{e^{ - {\rm{j}}\frac{{2\pi }}{\lambda }{\varphi _k}\left( {{{{\bf{\bar q}}}_{l,M}}} \right)}}} \right]^T},
\end{aligned}
 \end{equation}
 where ${r_{k,l}} = \left\| {{{\bf{\bar q}}_{l}} - {{\bf{w}}_k}} \right\|$ denotes the link distance between ${{{\bf{\bar q}}_{l}}}$ and ${{{\bf{w}}_k}}$. By applying the second-order Taylor approximation, we have
 \begin{equation}
 \begin{aligned}
 {r_{k,l}} &\approx r_{k,l}^{{\rm{second}}} \triangleq {r_k} + {x_l}{\Phi _k} + {y_l}{\Psi _k} + {z_l}{\Theta _k} + \\
 &\ \ \ \frac{{x_l^2 + y_l^2 + z_l^2 - {{\left( {{x_l}{\Phi _k} + {y_l}{\Psi _k} + {z_l}{\Theta _k}} \right)}^2}}}{{2{r_k}}},
 \end{aligned}
 \end{equation}
 where ${r_{k}} = \left\| {{{\bf{w}}_k} - {{\bf{q}}_r}} \right\|$ denotes the distance between ${\bf q}_r$ and ${\bf w}_k$.
 
 Moreover, the $L$ UAVs cooperate to form an $LM$-element distributed antenna array, and thus the channel from UE $k$ to the UAV swarm enabled MA system is
 {\footnote[1]{The LoS channel can be further generalized to the multi-path channel.}}
 \begin{equation}\label{channelUEk}
 {{\bf{h}}_k}\left( {\left\{ {{{\bf{\bar q}}_{l,m}},\forall l,m} \right\}} \right) = {\alpha _k}{{\bf{a}}_k}\left( {\left\{ {{{\bf{\bar q}}_{l,m}},\forall l,m} \right\}} \right),
 \end{equation}
 where ${\alpha _k} = \sqrt {{\beta _0}} /{r_{k}}$, with ${\beta _0}$ being the channel power at the reference distance of ${d_0} = 1$ m. Besides, ${{\bf{a}}_k}\left( {\left\{ {{{\bf{\bar q}}_{l,m}} ,\forall l,m} \right\}} \right) \in {\mathbb C}^{LM \times 1}$ denotes the receive array response vector of UE $k$, given by
 \begin{equation}\label{arrayResponseVector}
 \begin{aligned}
 {{\bf{a}}_k}\left( {\left\{ {{{\bf{\bar q}}_{l,m}},\forall l,m} \right\}} \right) = \left[ {{\bf{a}}_{k,1}^T\left( {\left\{ {{{\bf{\bar q}}_{1,m}},\forall m} \right\}} \right), \cdots ,} \right.\\
 {\left. {{\bf{a}}_{k,L}^T\left( {\left\{ {{{\bf{\bar q}}_{L,m}},\forall m} \right\}} \right)} \right]^T}.
 \end{aligned}
 \end{equation}
 In the following, for brevity, we use the notations ${{\bf{h}}_k}$, ${{\bf{a}}_k}$ and ${{\bf{a}}_{k,l}}$ to replace ${{\bf{h}}_k}\left( {\left\{ {{{\bf{\bar q}}_{l,m}},\forall l,m} \right\}} \right)$, ${{\bf{a}}_k}\left( {\left\{ {{{\bf{\bar q}}_{l,m}},\forall l,m} \right\}} \right)$ and ${{\bf{a}}_{k,l}}\left( {\left\{ {{{\bf{\bar q}}_{l,m}},\forall m} \right\}} \right)$, respectively.

 Denote by ${s_k}$ the independent and identically distributed (i.i.d.) information-bearing symbol of UE $k$, with ${\mathbb E}[ {{{\left| {{s_k}} \right|}^2}} ] = 1$. To detect its symbol, the receive beamforming ${{\bf{v}}_k} \in {\mathbb C}^{LM \times 1}$ is applied at the UAV swarm enabled MA system, with $\left\| {{\bf{v}}_k} \right\| = 1$. Thus, the resulting signal of UE $k$ after applying the receive beamforming is
 \begin{equation}\label{resultingSignalUEk}
 \begin{aligned}
 {y_k} = {\bf{v}}_k^H{{\bf{h}}_k}\sqrt {{P_k}} {s_k} +  {\bf{v}}_k^H\sum\limits_{i = 1,i \ne k}^K {{{\bf{h}}_i} \sqrt {{P_i}} {s_i}}  + {\bf{v}}_k^H{\bf{z}},
 \end{aligned}
 \end{equation}
 where ${{P_i}}$, $i = 1, \cdots, K$, denotes the transmit power of UE $i$, ${\bf{z}} \sim {\cal CN}\left( {{\bf{0}},{\sigma ^2}{{\bf{I}}_{LM}}} \right)$, with each element denoting the i.i.d. additive white Gaussian noise (AWGN) with zero mean and power ${\sigma ^2}$. To reveal the fundamental performance limit, synchronization is assumed for the UAV swarm system, which can be achieved via a synchronization protocol as in \cite{mohanti2019airbeam}.

 The resulting SINR for UE $k$ is
 \begin{equation}\label{SINRUEk}
 {\gamma _k} = \frac{{{{\bar P}_k}{{\left| {{\bf{v}}_k^H{{\bf{h}}_k}} \right|}^2}}}{{\sum\limits_{i = 1,i \ne k}^K {{{\bar P}_i}{{\left| {{\bf{v}}_k^H{{\bf{h}}_i}} \right|}^2}}  + 1}}= \frac{{{{\bar P}_k}{\bf{v}}_k^H{{\bf{h}}_k}{\bf{h}}_k^H{{\bf{v}}_k}}}{{{\bf{v}}_k^H{{\bf{C}}_k}{{\bf{v}}_k}}},
 \end{equation}
 where ${{\bf{C}}_k} \triangleq {{\bf{I}}_{LM}} + \sum\nolimits_{i = 1,i \ne k}^K {{{\bar P}_i}{{\bf{h}}_i}{\bf{h}}_i^H} $ denotes the interference-plus-noise covariance matrix with respect to (w.r.t.) UE $k$, and ${{\bar P}_i} \triangleq {P_i}/{\sigma ^2}$. The achievable rate of UE $k$  is
 \begin{equation}\label{achievableRateUEk}
 {R_k} = {\log _2}\left( {1 + {\gamma _k}} \right).
 \end{equation}

 Our objective is to maximize the minimum achievable rate over $K$ UEs, by jointly optimizing the 3D UAV swarm (relative) placement positions ${\bf{Q}} = \left\{ {{{\bf{q}}_l},\forall l} \right\}$, the local positions of MAs at all UAVs ${\bf{B}} = \left\{ {{b _{l,m}},\forall l,m = 2, \cdots ,M} \right\}$, and receive beamforming for all UEs ${\bf{V}} = \left\{ {{{\bf{v}}_k},\forall k} \right\}$. Let $\rho \left( {{\bf{Q}},{\bf{B}},{\bf{V}}} \right) = \mathop {\min }\limits_{k \in {\cal K}} {R_k}$, which is a function of ${\bf{Q}}$, ${\bf{B}}$, and ${\bf{V}}$. The optimization problem can be formulated as 
 \begin{align}\label{OptimizationProblem}
 \left( {\rm{P1}} \right)\ \ &\mathop {\max }\limits_{{\bf{Q}}, {\bf{B}},{\bf{V}}, \rho}\ \rho\\ 
 {\rm{s.t.}}&\ {R_k} \ge \rho,\ \forall k, \tag{\ref{OptimizationProblem}a}\\
 &\ {{\bf{q}}_r} + {{\bf{q}}_l} \in {\cal C},\ \forall l, \tag{\ref{OptimizationProblem}b}\\
 &\ {\left\| {{{\bf{q}}_l} - {{\bf{q}}_{l'}}} \right\|^2} \ge d_{\min }^2,\ \forall l, l' \ne l,\tag{\ref{OptimizationProblem}c}\\
 &\ {b_{l,m}} \in {\tilde {\cal C}},\ \forall l,m,\tag{\ref{OptimizationProblem}d}\\
 &\ {\left| {{b_{l,m}} - {b_{l,m'}}} \right|^2} \ge \tilde d_{\min }^2,\ \forall l,m,m' \ne m,\tag{\ref{OptimizationProblem}e}\\
 &\ \left\| {{{\bf{v}}_k}} \right\| = 1,\ \forall k, \tag{\ref{OptimizationProblem}f}
 \end{align}
 where $\cal C$ denotes the movable region of UAVs, ${\tilde{\cal C}}$ denotes the (local) movable region of MAs at each UAV, and the minimum safe distance constraint and mutual coupling avoidance constraint in \eqref{collisionAvoidanceConstraint} and \eqref{MCAvoidanceConstraint} have been replaced by their equivalent quadratic forms, i.e., (\ref{OptimizationProblem}c) and (\ref{OptimizationProblem}e), respectively. Problem (P1) is challenging to be directly solved due to the non-concave objective function and (some) non-convex constraints. Besides, the optimization variables ${\bf Q}$, ${\bf B}$ and ${\bf V}$ are intricately coupled in the objective function, which makes their joint optimization a challenging task.

\section{Special Case of Single-Antenna UAV}\label{sectionSingleAntennaUAV}
 In this section, to gain useful insights, we first consider the special case of single-antenna UAV, i.e., $M=1$, where (P1) is reduced to 
 \begin{equation}\label{OptimizationProblemSingleAntenna}
 \begin{aligned}
 \mathop {\max }\limits_{{\bf{Q}},{\bf{V}},\rho}&\ \rho\\
 {\rm{s.t.}}&\ {R_k} \ge \rho,\ \forall k,\\
 &\ {{\bf{q}}_r} + {{\bf{q}}_l} \in {\cal C},\ \forall l,\\
 &\ {\left\| {{{\bf{q}}_l} - {{\bf{q}}_{l'}}} \right\|^2} \ge d_{\min }^2,\ \forall l, l' \ne l,\\
 &\ \left\| {{{\bf{v}}_k}} \right\| = 1,\ \forall k.
 \end{aligned}
 \end{equation}
 In the following, the cases of one single UE, two UEs and arbitrary number of UEs are studied, respectively. 
 
 \subsection{Single UE and Two UEs}
 For the single UE communication, the receive signal in \eqref{resultingSignalUEk} reduces to
 \begin{equation}\label{resultingSignalSingleUE}
 y = {{\bf{v}}^H}{\bf{h}}\sqrt P s + {{\bf{v}}^H}{\bf{z}},
 \end{equation}
 where the UE index is omitted for brevity. By applying the optimal MRC beamforming, i.e., ${\bf{v}} = {\bf{h}}/\left\| {{\bf{h}}} \right\|$, the resulting SNR is given by
 \begin{equation}\label{SNRSingleUE}
 \gamma  = \bar P{\left\| {{\bf{h}}} \right\|^2} = \bar P{\left| \alpha  \right|^2}L,
 \end{equation}
 which only depends on the number of UAVs $L$, irrespective of the geometry formed by the UAV swarm enabled MA system. As a result, any UAV swarm placement within the movable region while satisfying the minimum safe distance constraint can achieve the maximum achievable rate.

 Next, we consider the classic MRC, zero-forcing (ZF) and minimum mean-square error (MMSE) beamforming schemes for the case of two UEs. Denote by ${\xi _{k,k'}} \triangleq \frac{{{{\left| {{\bf{h}}_k^H{{\bf{h}}_{k'}}} \right|}^2}}}{{{{\left\| {{{\bf{h}}_k}} \right\|}^2}{{\left\| {{{\bf{h}}_{k'}}} \right\|}^2}}}$  the channel's squared-correlation coefficient between UEs $k$ and $k'$, where $k,k'=1,2$ and $k \ne k'$. Besides, we have ${\left\| {{{\bf{h}}_k}} \right\|^2} = {\left| {{\alpha _k}} \right|^2}L$. The resulting SINR/SNR of UE $k$ after applying the three beamforming schemes are respectively given by \cite{lu2022near,lu2024tutorial}
 \begin{equation}\label{MRCSNRTwoUE}
 {\gamma _{{\rm{MRC}},k}} = {{\bar P}_k}{\left| {{\alpha _k}} \right|^2}L\left( {1 - \frac{{{{\bar P}_{k'}}{{\left| {{\alpha _{k'}}} \right|}^2}L{\xi _{k,k'}}}}{{{{\bar P}_{k'}}{{\left| {{\alpha _{k'}}} \right|}^2}L{\xi _{k,k'}} + 1}}} \right),
 \end{equation}
 \begin{equation}\label{ZFSNRTwoUE}
 {\gamma _{{\rm{ZF}},k}} = {{\bar P}_k}{\left| {{\alpha _k}} \right|^2}L\left( {1 - {\xi _{k,k'}}} \right),
 \end{equation}
 \begin{equation}\label{MMSESINRTwoUE}
 {\gamma _{{\rm{MMSE}},k}} = {{\bar P}_k}{\left| {{\alpha _k}} \right|^2}L\left( {1 - \frac{{{{\bar P}_{k'}}{{\left| {{\alpha _{k'}}} \right|}^2}L{\xi _{k,k'}}}}{{{{\bar P}_{k'}}{{\left| {{\alpha _{k'}}} \right|}^2}L + 1}}} \right).
\end{equation}
 It is observed from \eqref{MRCSNRTwoUE}-\eqref{MMSESINRTwoUE} that the SINRs of MRC, ZF, and MMSE beamforming depend on the coefficient ${\xi _{k,k'}}$, and a larger SINR can be achieved as ${\xi _{k,k'}}$ decreases. Moreover, the terms ${\frac{{{{\bar P}_{k'}}{{\left| {{\alpha _{k'}}} \right|}^2}L{\xi _{k,k'}}}}{{{{\bar P}_{k'}}{{\left| {{\alpha _{k'}}} \right|}^2}L{\xi _{k,k'}} + 1}}}$, ${\xi _{k,k'}}$, and ${\frac{{{{\bar P}_{k'}}{{\left| {{\alpha _{k'}}} \right|}^2}L{\xi _{k,k'}}}}{{{{\bar P}_{k'}}{{\left| {{\alpha _{k'}}} \right|}^2}L + 1}}}$ in \eqref{MRCSNRTwoUE}-\eqref{MMSESINRTwoUE} account for the SNR loss factors for UE $k$ due to applying the MRC, ZF, and MMSE beamforming schemes, respectively. By substituting \eqref{channelUEk} into ${\xi _{k,k'}}$, we have
 \begin{equation}
 {\xi _{k,k'}} = \frac{1}{{{L^2}}}{\left| {{\bf{a}}_k^H{{\bf{a}}_{k'}}} \right|^2}.
 \end{equation}

 Based on the above observation, for the case of two UEs, we aim to minimize the channels' squared-correlation coefficient, which can be formulated as 
 \begin{equation}\label{OptimizationProblemTwoUE}
 \begin{aligned}
 \mathop {\min }\limits_{\bf{Q}} &\  \frac{1}{{{L^2}}}{\left| {{\bf{a}}_k^H{{\bf{a}}_{k'}}} \right|^2}\\
 {\rm{s.t.}}&\ {{\bf{q}}_r} + {{\bf{q}}_l} \in {\cal C},\ \forall l, \\
 &\ {\left\| {{{\bf{q}}_l} - {{\bf{q}}_{l'}}} \right\|^2} \ge d_{\min }^2,\ \forall l, l' \ne l.
 \end{aligned}
 \end{equation}

 To tackle problem \eqref{OptimizationProblemTwoUE}, we first study the property of the objective function. Note that when the movable region size is much smaller than the link distance, the UPW model gives a valid approximation of spherical wavefront for channels between UAVs and ground UEs \cite{lu2024tutorial}, which may correspond to the cases of small movable region and/or high UAV operation altitude. In this case, the receive response vector in \eqref{arrayResponseVector} reduces to 
 ${{\bf{a}}_k} = {e^{ - {\rm{j}}\frac{{2\pi }}{\lambda }{r_k}}}{[ {{e^{ - {\rm{j}}\frac{{2\pi }}{\lambda }{\bm{\kappa }}_k^T{{\bf{q}}_1} } }, \cdots ,{e^{ - {\rm{j}}\frac{{2\pi }}{\lambda }{\bm{\kappa }}_k^T{{\bf{q}}_L} }}}]^T}$. By substituting it into the objective function, we have
 \begin{equation}\label{correlationCoefficient}
 \frac{1}{{{L^2}}}{\left| {{\bf{a}}_k^H{{\bf{a}}_{k'}}} \right|^2} = \frac{1}{{{L^2}}}{\left| {\sum\limits_{l = 1}^L {{e^{  {\rm{j}}\frac{{2\pi }}{\lambda }{{\left( {{{\bm{\kappa }}_k} - {{\bm{\kappa }}_{k'}}} \right)}^T}{{\bf{q}}_l}}}} } \right|^2}.
 \end{equation}

 By letting ${\nu _{\min }} =  {\left\lceil {L{d_{\min }}\left\| {{{\bm{\kappa }}_k} - {{\bm{\kappa }}_{k'}}} \right\|/\lambda  - 1} \right\rceil } $, we obtain the following theorem for UAV position without drift. In particular, high-accuracy positioning technologies, such as real-time kinematic (RTK) global navigation satellite system (GNSS), multi-sensor fusion, and robust flight control algorithms help to mitigate the issue of UAV position drift \cite{zuo2022unmanned}.

 \begin{theorem} \label{zeroInnerProductTheorem}
 An optimal solution to \eqref{OptimizationProblemTwoUE} that achieves zero objective value is
 \begin{equation}\label{zeroInnerProductTrajectory}
 {{\bf{q}}_l} ={{\bf{q}}_1} + \left( {l - 1} \right)\frac{{\left( {{\varsigma^{\star}}  + 1/L} \right)\lambda }}{{{{\left\| {{{\bm{\kappa }}_k} - {{\bm{\kappa }}_{k'}}} \right\|}^2}}}\left( {{{\bm{\kappa }}_k} - {{\bm{\kappa }}_{k'}}} \right),
 \end{equation}
 where ${{\bf{q}}_1}$ can be any vector to guarantee ${{\bf{q}}_r} + {{\bf{q}}_l} \in {\cal C}$, $\forall l$, and ${\varsigma^{\star}}$ is given by 
 \begin{equation}\label{minimumInteger}
 {\varsigma^{\star}}  = \left\{ \begin{split}
 &\frac{{{\nu _{\min }}}}{L},\ \ \ \ \ \ {\rm if}\ \bmod\left( {{\nu _{\min }} + 1,L} \right) \ne 0,\\
 &\frac{{{\nu _{\min }} + 1}}{L},\ {\rm otherwise}.
 \end{split} \right.
 \end{equation}
 \end{theorem}
 \begin{IEEEproof}
 Please refer to Appendix~\ref{proofOfzeroInnerProductTheorem}.
 \end{IEEEproof}
 
 Note that Theorem~\ref{zeroInnerProductTheorem} extends the 1D result in \cite{zhu2023full} to the general 3D placement, and the derived closed-form solution ${\varsigma^{\star}}$ can be a fraction, which removes the requirement of being an integer imposed in \cite{zhu2023full}. Moreover, a sufficient condition for ${{\bf{q}}_r} + {{\bf{q}}_l} \in {\cal C}$, $\forall l$, is that ${\cal C}$ contains a line segment parallel to $({\bm{\kappa}}_k-{\bm{\kappa}}_{k'})$ with length no smaller than $\left\| {{{\bf{q}}_L} - {{\bf{q}}_1}} \right\| = \frac{{\left( {L - 1} \right)\left( {{\varsigma ^ \star } + 1/L} \right)\lambda }}{{\left\| {{{\bm{\kappa }}_k} - {{\bm{\kappa }}_{k'}}} \right\|}}$.
 
 With \eqref{zeroInnerProductTrajectory} and \eqref{minimumInteger}, we have ${\xi _{k,k'}} = 0$, and ${\gamma _{{\rm{MRC}},k}} = {\gamma _{{\rm{ZF}},k}}\ = {\gamma _{{\rm{MMSE}},k}} = {{\bar P}_k}{\left\| {{{\bf{h}}_k}} \right\|^2} = {{\bar P}_k}{\left| {{\alpha _k}} \right|^2}L$, i.e., they are identical to the single UE SNR without IUI. Thus, an IUI-free communication can be achieved with the UAV swarm enabled MA system, by setting the UAVs' static positions according to Theorem~\ref{zeroInnerProductTheorem}. It is worth noting that the adjacent inter-UAV distance is $\frac{{\left| {{\varsigma ^ {\star} } + 1/L} \right|\lambda }}{{\left\| {{{\bm{\kappa }}_k} - {{\bm{\kappa }}_{k'}}} \right\|}} \ge d_{\min}$, where the minimum distance $d_{\min}$ is in general much larger than half wavelength. Thus, the UAV swarm forms a USA for the IUI-free communication. In particular, compared to the classic compact array where antenna spacing is separated by half wavelength, USA can achieve a larger array aperture, and thus a narrower beamwidth of the main lobe is enabled.
 
 Moreover, the objective function in (P1) is given by 
 \begin{equation}\label{minrateTwoUESingleAntenna}
 \mathop {\min }\limits_{k \in {\cal K}} {\log _2}\left( {1 + {{\bar P}_k}{{\left| {{\alpha _k}} \right|}^2}L} \right).
 \end{equation}
 
 It can be observed that the UAV swarm enabled MA system not only completely eliminates the IUI, but also achieves the full beamforming gain at each UE, in terms of the number of UAVs $L$ in \eqref{minrateTwoUESingleAntenna}, without suffering from any performance loss as in the traditional FPA system \cite{lu2022near}.

 \subsection{Arbitrary Number of UEs}\label{subSectiongeneralUEsSingleAntenna}
 In this subsection, for arbitrary number of UEs, we propose an alternating optimization algorithm to solve problem \eqref{OptimizationProblemSingleAntenna} sub-optimally, where the receive beamforming and each UAV placement position are optimized alternately in an iterative manner. 
 
 \subsubsection{Optimization of ${\bf V}$ With Given ${\bf Q}$}
 For given UAV swarm placement positions ${\bf Q}$, the receive beamforming optimization problem in \eqref{OptimizationProblemSingleAntenna} is expressed as
 \begin{equation}\label{subProblemBeamforming}
 \begin{aligned}
 \mathop {\max }\limits_{\bf{V},\rho} &\ \rho \\
 {\rm{s.t.}}&\ {R_k} \ge \rho,\ \forall k,\\
 &\ \left\| {{{\bf{v}}_k}} \right\| = 1,\ \forall k.
 \end{aligned}
 \end{equation}

 A closer look at \eqref{SINRUEk} shows that the receive beamforming ${{\bf{v}}_k}$ only impacts the SINR for UE $k$, and \eqref{SINRUEk} is a generalized Rayleigh quotient w.r.t. ${\bf{v}}_k$, the optimal ${\bf{v}}_k$ to \eqref{subProblemBeamforming} is given by the classical MMSE beamforming \cite{wang2025optimal}:
 \begin{equation}\label{optimalReceiveBeamforming}
 {\bf{v}}_k^ \star  = \frac{{{\bf{C}}_k^{ - 1}{{\bf{h}}_k}}}{{\left\| {{\bf{C}}_k^{ - 1}{{\bf{h}}_k}} \right\|}},\ \forall k.
 \end{equation}

 \subsubsection{Optimization of ${{\bf{q}}_l}$ With Given ${\bf V}$ and $\left\{ {{\bf{q}}_{l'} ,\forall l' \ne l} \right\}$}\label{subsubsectionUAVTrajectory}
 For given ${\bf V}$ and $\left\{ {{\bf{q}}_{l'},\forall l' \ne l} \right\}$, the sub-problem of \eqref{OptimizationProblemSingleAntenna} for optimizing the placement position of UAV $l$ can be expressed as  
 \begin{align}\label{subProblemTrajectory1}
 \mathop {\max }\limits_{{\bf{q}}_l,\rho }  &\ \rho  \\
 {\rm{s.t.}}&\ {R_k}  \ge \rho,\ \forall k \in {\cal K},\tag{\ref{subProblemTrajectory1}a}\\
 &\ {{\bf{q}}_r} + {{\bf{q}}_l} \in {\cal C},\tag{\ref{subProblemTrajectory1}b}\\
 &\ {\left\| {{{\bf{q}}_l} - {{\bf{q}}_{l'}}} \right\|^2} \ge d_{\min }^2,\ \forall l' \ne l. \tag{\ref{subProblemTrajectory1}c}
 \end{align}

 Moreover, to tackle the non-convexity of the constraint (\ref{subProblemTrajectory1}a), the slack variables $\left\{ {{\eta_k},{\mu_k},\forall k} \right\}$ are introduced such that
 \begin{equation}
 {e^{{\eta_k}}} = 1 + \sum\limits_{i = 1}^K {{{\bar P}_i}{{\left| {{\bf{v}}_k^H{{\bf{h}}_i}} \right|}^2}},
 \end{equation}
 \begin{equation}
 {e^{{\mu _k}}} = 1 + \sum\limits_{i = 1,i \ne k}^K {{{\bar P}_i}{{\left| {{\bf{v}}_k^H{{\bf{h}}_i}} \right|}^2}}.
 \end{equation}

 Thus, problem \eqref{subProblemTrajectory1} is transformed into
 \begin{align}\label{subProblemTrajectory2}
 &\mathop {\max }\limits_{{{\bf q}_l},\rho,\left\{ {{\eta _k},{\mu _k}} \right\}_{k=1}^K} \ \ \rho \\
 {\rm{s.t.}}&\  {\frac{{\ln \left( {{e^{{\eta _k} - {\mu _k}}}} \right)}}{{\ln 2}}} \ge \rho,\ \forall k \in {\cal K},\tag{\ref{subProblemTrajectory2}a}\\
 & {\sigma ^2} + \sum\limits_{i = 1}^K {{P_i}{{\left| {{\bf{v}}_k^H{{\bf{h}}_i}} \right|}^2}}  \ge {\sigma ^2}{e^{{\eta _k}}},\ \forall k \in {\cal K},\tag{\ref{subProblemTrajectory2}b}\\
 & {\sigma ^2} + \sum\limits_{i = 1,i \ne k}^K {{P_i}{{\left| {{\bf{v}}_k^H{{\bf{h}}_i}} \right|}^2}}  \le {\sigma ^2}{e^{{\mu _k}}},\ \forall k \in {\cal K},\tag{\ref{subProblemTrajectory2}c}\\
 & {{\bf{q}}_r} + {{\bf{q}}_l} \in {\cal C},\tag{\ref{subProblemTrajectory2}d}\\
 & {\left\| {{{\bf{q}}_l} - {{\bf{q}}_{l'}}} \right\|^2} \ge d_{\min }^2,\ l' \ne l.\tag{\ref{subProblemTrajectory2}e}
 \end{align}
 
 Note that problem \eqref{subProblemTrajectory2} is still challenging to be solved since the constraints (\ref{subProblemTrajectory2}b), (\ref{subProblemTrajectory2}c) and  (\ref{subProblemTrajectory2}e) are non-convex. To tackle this issue, the SCA technique \cite{zeng2019accessing} is applied in the following, which is an efficient iterative optimization technique that successively updates the optimization variables by solving the approximated convex problem at each iteration.

 Let ${{\bf{\bar h}}_{i,l}} \in {{\mathbb C}^{\left( {L - 1} \right) \times 1}}$ denote the resulting channel after removing ${h_{i,l}}$ from ${{\bf{h}}_i}$, and ${{\bf{\bar v}}_{k,l}}  \in {{\mathbb C}^{\left( {L - 1} \right) \times 1}}$ denote the resulting beamforming vector after removing ${v_{k,l}}$ from ${{\bf{v}}_k}$, where ${h_{i,l}}$ and ${v_{k,l}}$ denote the $l$-th element of ${{\bf{h}}_i}$ and ${{\bf{v}}_k}$, respectively. Then, ${f_{k,i}} \triangleq {\left| {{\bf{v}}_k^H{{\bf{h}}_i}} \right|^2}$ can be expressed as 
 \begin{equation}\label{expressionfki}
 \begin{aligned}
 {f_{k,i}} &= {\bf{h}}_i^H{{\bf{v}}_k}{\bf{v}}_k^H{{\bf{h}}_i} =  \underbrace {\sum\limits_{l' = 1,l' \ne l}^L {2{\mathop{\rm Re}\nolimits} \left\{ {h_{i,l}^ * {V_{k,l,l'}}{h_{i,l'}}} \right\}} }_{{g_{k,i,l}}}  \\
 &+ \underbrace {{{\left| {{\alpha _i}} \right|}^2}{V_{k,l,l}} + {\bf{\bar h}}_{i,l}^H{{\bf{\bar V}}_{k,l,l}}{{\bf{\bar h}}_{i,l}}}_{{{\bar g}_{k,i,l}}},
 \end{aligned}
 \end{equation}
 where ${V_{k,l,l'}} \triangleq {v_{k,l}}v_{k,l'}^ * $, ${{\bf{\bar V}}_{k,l,l}} \in {{\mathbb C}^{\left( {L - 1} \right) \times \left( {L - 1} \right)}} \triangleq {{\bf{\bar v}}_{k,l}}{\bf{\bar v}}_{k,l}^H$, and ${\bar g}_{k,i,l}$ is independent of $ {{\bf{q}}_l} $. By substituting ${h_{i,l}} = {\alpha _i}{e^{ - {\rm{j}}\frac{{2\pi }}{\lambda }{r_{i,l}}}}$ into ${g_{k,i,l}}$, we have 
 \begin{equation}\label{expressiongkim}
 \begin{aligned}
 &{g_{k,i,l}} = \sum\limits_{l' = 1,l' \ne l}^L {2{{\left| {{\alpha _i}} \right|}^2}{\rm{Re}}\left\{ {{V_{k,l,l'}}{e^{{\rm{j}}\frac{{2\pi }}{\lambda }\left( {{r_{i,l}} - {r_{i,l'}}} \right)}}} \right\}} \\
 &= \sum\limits_{l' = 1,l' \ne l}^L {2{{\left| {{\alpha _i}} \right|}^2}\left| {{V_{k,l,l'}}} \right|} \cos \left( {\frac{{2\pi }}{\lambda }\left( {{r_{i,l}} - {r_{i,l'}}} \right) + \angle {V_{k,l,l'}}} \right).
 \end{aligned}
 \end{equation}

 It is observed that \eqref{expressiongkim} is neither convex nor concave w.r.t. ${{\bf q}_l}$. To this end, two quadratic surrogate functions are respectively constructed to serve as the lower and upper bounds for ${g_{k,i,l}}$, shown in the following lemma.  
 
 \begin{lemma} \label{quadraticSurrogatelemma}
 The lower and upper bounds of ${g_{k,i,l}}$ are respectively given by 
 \begin{equation}\label{lowerBoundgkim}
 \begin{aligned}
 {g_{k,i,l}} &\ge {\left[ {{g_{k,i,l}}} \right]_{\rm lb}} = {g_{k,i,l}}\left( {{\bf{q}}_l^{\left( j \right)}} \right) + \nabla {g_{k,i,l}}{\left( {{\bf{q}}_l^{\left( j \right)}} \right)^T}\left( {{{\bf{q}}_l} - {\bf{q}}_l^{\left( j \right)}} \right)\\
 &\ \ \ \ \ \ \ \ - \frac{{{\delta _{k,i,l}}}}{2}{\left( {{{\bf{q}}_l} - {\bf{q}}_l^{\left( j \right)}} \right)^T}\left( {{{\bf{q}}_l} - {\bf{q}}_l^{\left( j \right)}} \right),
 \end{aligned}
 \end{equation}
 \begin{equation}\label{upperBoundgkim}
 \begin{aligned}
 {g_{k,i,l}} &\le {\left[ {{g_{k,i,l}}} \right]_{\rm ub}} = {g_{k,i,l}}\left( {{\bf{q}}_l^{\left( j \right)}} \right) + \nabla {g_{k,i,l}}{\left( {{\bf{q}}_l^{\left( j \right)}} \right)^T}\left( {{{\bf{q}}_l} - {\bf{q}}_l^{\left( j \right)}} \right)\\
 &\ \ \ \ \ \ \ \ + \frac{{{\delta _{k,i,l}}}}{2}{\left( {{{\bf{q}}_l} - {\bf{q}}_l^{\left( j \right)}} \right)^T}\left( {{{\bf{q}}_l} - {\bf{q}}_l^{\left( j \right)}} \right),
 \end{aligned}
 \end{equation}
 where ${\delta _{k,i,l}} = \frac{{2\pi }}{\lambda }\sum\limits_{l' = 1,l' \ne l}^L {2{{\left| {{\alpha _i}} \right|}^2}\left| {{V_{k,l,l'}}} \right|} \sqrt {\frac{2}{{r_i^2}} + {{\left( {\frac{{2\pi }}{\lambda }} \right)}^2}{{\left( {1 + \frac{{{C_{\max }}}}{{r_i^2}}} \right)}^2}}$, ${{\bf{q}}_l^{\left( j \right)}}$ denotes the resulting placement position of UAV $l$ in the $j$-th iteration, and $\nabla {g_{k,i,l}}( {{\bf{q}}_l^{\left( j \right)}} )$ denotes the gradient of ${{g_{k,i,l}}}$ over ${\bf{q}}_l^{\left( j \right)}$.
 \end{lemma}

 \begin{IEEEproof}
 Please refer to Appendix \ref{proofOfquadraticSurrogatelemma}.
 \end{IEEEproof}
 
 With Lemma \ref{quadraticSurrogatelemma}, the lower and upper bounds of the function ${f_{k,i}}$ are 
 \begin{equation}
 {f_{k,i}} \ge {\left[ {{f_{k,i}}} \right]_{\rm lb}} \triangleq {\left[ {{g_{k,i,l}}} \right]_{\rm lb}} + {{\bar g}_{k,i,l}},
 \end{equation}
 \begin{equation}
 {f_{k,i}} \le {\left[ {{f_{k,i}}} \right]_{\rm ub}} \triangleq {\left[ {{g_{k,i,l}}} \right]_{\rm ub}} + {{\bar g}_{k,i,l}},
 \end{equation}
 which are concave and convex w.r.t. ${{\bf{q}}_l}$, respectively.  
 
 Moreover, to tackle the non-convexity of ${e^{{\mu _k}}}$ on the right-hand-side (RHS) of (\ref{subProblemTrajectory2}c), let $\mu _k^{\left( j \right)}$ denote the resulting variable in the $j$-th iteration; then the convex function ${e^{{\mu _k}}}$ is lower-bounded by
 \begin{equation}
 {e^{{\mu _k}}} \ge {\left[ {{e^{{\mu _k}}}} \right]_{{\rm{lb}}}} \triangleq {\mu _k}{e^{\mu _k^{\left( j \right)}}} + \left( {1 - \mu _k^{\left( j \right)}} \right){e^{\mu _k^{\left( j \right)}}}.
 \end{equation}

 Similarly, for given ${\bf{q}}_l^{\left( j \right)}$, the convex function ${\left\| {{{\bf{q}}_l} - {{\bf{q}}_{l'}}} \right\|^2}$ on the left-hand-side (LHS) of (\ref{subProblemTrajectory2}e) is lower-bounded by \cite{zeng2019accessing}
 \begin{equation}
 \begin{aligned}
 &{\left\| {{{\bf{q}}_l} - {{\bf{q}}_{l'}}} \right\|^2} \ge {\left[ {{{\left\| {{{\bf{q}}_l} - {{\bf{q}}_{l'}}} \right\|}^2}} \right]_{{\rm{lb}}}}\\
 &\triangleq  2{\left( {{\bf{q}}_l^{\left( j \right)} - {{\bf{q}}_{l'}}} \right)^T}\left( {{{\bf{q}}_l} - {\bf{q}}_l^{\left( j \right)}} \right)  + {\left\| {{\bf{q}}_l^{\left( j \right)} - {{\bf{q}}_{l'}}} \right\|^2}.
 \end{aligned}
 \end{equation}

 As a result, problem \eqref{subProblemTrajectory2} is lower-bounded by the following problem for given $\{ {{\bf{q}}_l^{\left( j \right)},\mu _k^{\left( j \right)}} \}$.
 \begin{equation}\label{subProblemTrajectory3}
 \begin{aligned}
 &\mathop {\max }\limits_{{{\bf{q}}_l},\rho,\left\{ {{\eta _k},{\mu _k}} \right\}_{k=1}^K} \ \ \rho \\
 {\rm{s.t.}}&\  {{\eta _k} - {\mu _k}}   \ge \rho \ln 2,\ \forall k \in {\cal K},\\
 & {\sigma ^2} + \sum\limits_{i = 1}^K {{P_i}{{\left[ {{f_{ki}}} \right]}_{{\rm{lb}}}}}  \ge {\sigma ^2}{e^{{\eta _k}}},\ \forall k \in {\cal K},\\
 & {\sigma ^2} + \sum\limits_{i = 1,i \ne k}^K {{P_i}{{\left[ {{f_{ki}}} \right]}_{{\rm{ub}}}}}  \le {\sigma ^2}{\left[ {{e^{{\mu _k}}}} \right]_{{\rm{lb}}}},\ \forall k \in {\cal K},\\
 & {{\bf{q}}_r} + {{\bf{q}}_l} \in {\cal C},\\
 &{\left[ {{{\left\| {{{\bf{q}}_l} - {{\bf{q}}_{l'}}} \right\|}^2}} \right]_{{\rm{lb}}}} \ge d_{\min }^2,\ \forall l' \ne l,
 \end{aligned}
 \end{equation}
 which is a convex optimization problem and can be solved via the standard convex optimization tools, such as CVX.

 The overall algorithm is summarized in Algorithm~\ref{alg1}, and its convergence proof is given as follows.

 \begin{proposition}\label{Convergence}
 The objective value of problem \eqref{OptimizationProblemSingleAntenna} obtained by Algorithm~\ref{alg1} is guaranteed to converge.
 \end{proposition}

 \begin{IEEEproof}
 Please refer to Appendix~\ref{proofOfSingleAntennaUAVConvergence}.
 \end{IEEEproof} 

 It is worth mentioning that the objective value at each iteration is non-decreasing, thus guaranteeing the convergence of Algorithm~\ref{alg1}. Moreover, the computational complexity is analyzed as follows. In step 3, the complexity for obtaining the receive beamforming is ${\cal O}( {K{L^3}} )$. The complexity from step 4 to step 6 is approximately ${\cal O}( {I_1}L{{( {2K})}^3} )$, where $I_1$ denotes the maximum number of iterations required by SCA for convergence. Thus, the total computational complexity of Algorithm~\ref{alg1} is ${\cal O}( {{I_2}K{L^3} + {I_1}{I_2}L{( {2K} )}^3} )$, where $I_2$ denotes the number of iterations required by the alternating optimization for convergence. 
 
 \begin{algorithm}[t]
 \caption{Proposed Alternating Optimization for Solving Problem \eqref{OptimizationProblemSingleAntenna}}
 \label{alg1}
 \begin{algorithmic}[1]
 \STATE Initialize ${{\bf{Q}}^{\left( 0 \right)}}$ and $\{ {{\bf{v}}_k^{\left( 0 \right)}} \}$ randomly, and let $j=0$.
 \REPEAT
 \STATE For given ${{\bf{Q}}^{\left( j \right)}}$, obtain the optimal solution to \eqref{subProblemBeamforming} based on \eqref{optimalReceiveBeamforming}, denoted as $\{ {{\bf{v}}_k^{\left( {j + 1} \right)}} \}$. 
 \STATE \textbf{for} $l = 1:L$ \textbf{do}
 \STATE \quad Obtain $\{ {\bf{q}}_l^{\left( {j + 1} \right)}\}$ by solving problem \eqref{subProblemTrajectory3}, given \\
 \quad $\{ {\bf{q}}_1^{\left( {j + 1} \right)}, \cdots ,{\bf{q}}_{l - 1}^{\left( {j + 1} \right)},{\bf{q}}_l^{\left( j \right)}, \cdots ,{\bf{q}}_L^{\left( j \right)}\} $ and $\{ {{\bf{v}}_k^{\left( {j + 1} \right)}} \}$.
 \STATE \textbf{end for}
 \STATE Update $j=j+1$.
 \UNTIL the fractional increase in the objective function value is below a given threshold $ \epsilon > 0$.
 \end{algorithmic}
 \end{algorithm}

\section{Multi-Antenna UAV}\label{sectionMultiAntennaUAV}
 In this section, we consider the general case of multi-antenna UAV, where each UAV deploys an MA array and all the UAVs simultaneously forms a larger-scale MA system with their individual MA arrays.

\subsection{Single UE and Two UEs}
 For the single UE communication, similar to \eqref{SNRSingleUE}, the resulting SNR is 
 \begin{equation}\label{SNRSingleUEMultiAntenna}
 \gamma  = \bar P{\left\| {{\bf{h}}} \right\|^2} = \bar P{\left| \alpha  \right|^2}LM.
 \end{equation}
 
 For the scenario with two UEs, the resulting SINR/SNR of UE $k$ with the MRC, ZF, and MMSE beamforming can be similarly obtained by substituting $L$ with $LM$ in \eqref{MRCSNRTwoUE}-\eqref{MMSESINRTwoUE}, respectively. Besides, when UPW model holds, the channel's squared-correlation coefficient between UEs $k$ and $k'$ is
 \begin{equation}\label{correlationCoefficientMultiAntenna}
 \begin{aligned}
 {\xi _{k,k'}} &= \frac{1}{{{L^2}{M^2}}}{\left| {\sum\limits_{l = 1}^L {{\bf{a}}_{k,l}^H{{\bf{a}}_{k',l}}} } \right|^2}\\
 & = \frac{1}{{{L^2}{M^2}}}{\left| {\sum\limits_{l = 1}^L {{e^{ {\rm{j}}\frac{{2\pi }}{\lambda }\left( {{\bm{\kappa }}_k^T - {\bm{\kappa }}_{k'}^T} \right){{\bf{q}}_l}}}\sum\limits_{m = 1}^M {{e^{ {\rm{j}}\frac{{2\pi }}{\lambda }{b_{l,m}} \Delta {\Phi _{k,k'}} }}} } } \right|^2},
 \end{aligned}
 \end{equation}
 where $\Delta {\Phi _{k,k'}} \triangleq {\Phi _k} - {\Phi _{k'}}$. It is observed from \eqref{correlationCoefficientMultiAntenna} that when $\sum\nolimits_{m = 1}^M {{e^{  {\rm{j}}\frac{{2\pi }}{\lambda }{b_{l,m}}\Delta {\Phi _{k,k'}}}}}=0$, $\forall l$, we have ${\xi _{k,k'}} = 0$.

 Similar to Theorem~\ref{zeroInnerProductTheorem}, an optimal solution of $\left\{ {{b_{l,m}}} \right\}$ to \eqref{correlationCoefficientMultiAntenna} that achieves zero objective value is
 \begin{equation}\label{zeroInnerProductTrajectoryMultiAntenna}
 {b_{l,m}} = \left( {m - 1} \right)\frac{{\left( {\zeta  _l^{\star}  + 1/M} \right)\lambda }}{\left|{\Delta {\Phi _{k,k'}}}\right|},
 \end{equation}
 where the common coefficient ${\zeta  _l^{\star}}$ is derived to satisfy the mutual coupling avoidance constraint between adjacent MA elements for each UAV, given by 
 \begin{equation}\label{minimumIntegerMultiAntenna}
 {\zeta_l ^ \star } = \left\{ \begin{split}
 &\frac{{{{\tilde \nu }_{\min }}}}{M},\ \ \ \ \ \ {\rm if}\ \bmod\left( {{{\tilde \nu }_{\min }} + 1,M} \right) \ne 0,\\
 &\frac{{{{\tilde \nu }_{\min }} + 1}}{M},\ {\rm otherwise}.
 \end{split} \right.
 \end{equation}
 with ${{\tilde \nu }_{\min }} = \left\lceil {M{{\tilde d}_{\min }}\left| {\Delta {\Phi _{k,k'}}} \right|/\lambda  - 1} \right\rceil$ common to all UAVs. Moreover, MA position ${b_{l,m}} \in \tilde {\cal C}$, $\forall l,m$, can be satisfied when $\left| {{b_{l,M}} - {b_{l,1}}} \right| = \left( {M - 1} \right)\frac{{\left( {\zeta _l^ \star  + 1/M} \right)\lambda }}{{\left| {\Delta {\Phi _{k,k'}}} \right|}}$ is no greater than the maximum distance between any two points in ${\tilde {\cal C}}$.
 
 Note that the MAs' positions given in \eqref{zeroInnerProductTrajectoryMultiAntenna}-\eqref{minimumIntegerMultiAntenna} enable an IUI-free communication via MAs' positions adjustment, irrespective of UAV swarm placement positions. The objective function in (P1) is thus given by
 \begin{equation}\label{minrateTwoUEMultiAntenna}
 \mathop {\min }\limits_{k \in {\cal K}}  {{{\log }_2}\left( {1 + {{\bar P}_k}{{\left| {{\alpha _k}} \right|}^2}LM} \right)}.
 \end{equation}
 
 In fact, the results in \eqref{zeroInnerProductTrajectoryMultiAntenna}-\eqref{minimumIntegerMultiAntenna} correspond to the case where all MA arrays have the identical architecture. To gain further insights, we express ${{\bf{b}}_{l,m}}$ as  ${{\bf{b}}_{l,m}} = {{\bf{b}}_m} = {\left[ {{b_m},0,0} \right]^T}$, $\forall l$, the channels' squared-correlation coefficient in \eqref{correlationCoefficientMultiAntenna} is reduced to
 \begin{equation}\label{correlationCoefficientMultiAntennaSpecial}
 {\xi _{k,k'}} = \frac{1}{{{L^2}}}{\left| {\sum\limits_{l = 1}^L {{e^{ {\rm{j}}\frac{{2\pi }}{\lambda }\left( {{\bm{\kappa }}_k^T - {\bm{\kappa }}_{k'}^T} \right){{\bf{q}}_l}}}} } \right|^2}\underbrace {\frac{1}{{{M^2}}}{{\left| {\sum\limits_{m = 1}^M {{e^{ {\rm{j}}\frac{{2\pi }}{\lambda }{b_{m}}\Delta {\Phi _{k,k'}}}}} } \right|}^2}}_{ \le 1}.
 \end{equation}
 It is observed that \eqref{correlationCoefficientMultiAntennaSpecial} differs from \eqref{correlationCoefficient} in the term ${\frac{1}{{{M^2}}}{{| {\sum\nolimits_{m = 1}^M {{e^{  {\rm{j}}\frac{{2\pi }}{\lambda }{b_{m}}\Delta {\Phi _{k,k'}}}}} } |}^2}} \le 1$, which characterizes the capability of the MA array mounted on each UAV to distinguish UE $k$ and $k'$ in the spatial domain. Thus, the IUI can be mitigated by adjusting not only the UAV swarm placement positions, but also the local positions of MAs at all UAVs.  
 
 In particular, when the USA is adopted, i.e., ${b_m} = \left( {m - 1} \right)\Gamma d = \left( {m - 1} \right)\Gamma \lambda /2$, where $\Gamma \ge 1$ denotes the sparsity level \cite{wang2024enhancing,lu2024group}, \eqref{correlationCoefficientMultiAntennaSpecial} reduces to 
 \begin{equation}
 \begin{aligned}
 {\xi _{k,k'}} &= \frac{1}{{{L^2}}}{\left| {\sum\limits_{l = 1}^L {{e^{ {\rm{j}}\frac{{2\pi }}{\lambda }\left( {{\bm{\kappa }}_k^T - {\bm{\kappa }}_{k'}^T} \right){{\bf{q}}_l}}}} } \right|^2} \times \\
 &\ \ \ \ \ \ \ \ \ \ \ \ \ \ \ \ \  \ \ \ \ \ \frac{1}{{{M^2}}}\left| {\frac{{\sin \left( {\frac{{\pi M\Gamma \Delta {\Phi _{k,k'}}}}{2}} \right)}}{{\sin \left( {\frac{{\pi \Gamma \Delta {\Phi _{k,k'}}}}{2}} \right)}}} \right|^2.
 \end{aligned}
 \end{equation}

 Then, the coefficient ${\xi _{k,k'}}$ is equal to zero when sparsity level satisfies 
 \begin{equation}
 \Gamma  \in \left\{ {\left. {\frac{{2\varepsilon }}{{M\Delta {\Phi _{k,k'}}}}} \right|\varepsilon  \in {\mathbb Z},{\frac{{2\varepsilon }}{{M\Delta {\Phi _{k,k'}}}} \ge 1} ,\bmod \left( {\varepsilon ,M} \right) \ne 0} \right\}.
 \end{equation}

 \begin{figure}[!t]
 \centering
 \centerline{\includegraphics[width=3.5in,height=2.625in]{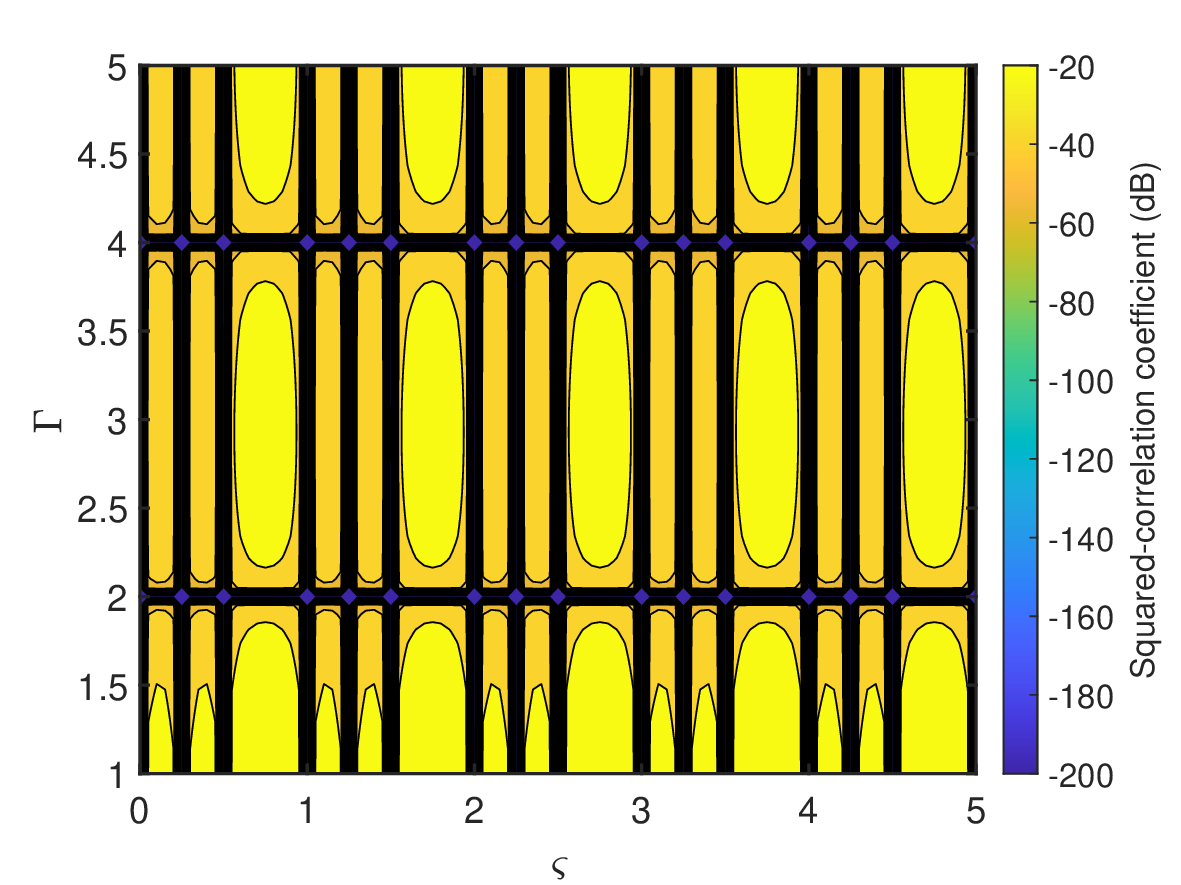}}
 \caption{The channels' squared-correlation coefficient ${\xi _{1,2}}$ versus $\varsigma$ and $\Gamma$.}
 \label{fig:correlationCoefficientVersusVarsigmaGamma}
 \end{figure}
 
 Fig.~\ref{fig:correlationCoefficientVersusVarsigmaGamma} shows the channels' squared-correlation coefficient ${\xi _{1,2}}$ versus $\varsigma$ and $\Gamma$ for the two-UE communication, by considering the placement positions given in \eqref{zeroInnerProductTrajectory} and \eqref{minimumInteger}. The UE directions are $\left( {{\theta _1},{\phi _1}} \right) = \left( {{{30}^ \circ },{{60}^ \circ }} \right)$ and $\left( {{\theta _2},{\phi _2}} \right) = \left( {{{0}^ \circ },{{0}^ \circ }} \right)$, respectively. The number of UAVs is $L = 4$, and each UAV is equipped with a USA with $M = 4$ antennas. For convenience of presentation, the coefficient ${\xi _{1,2}}$ below $-200$ dB is truncated to $-200$ dB. It is observed that when $\varsigma$ satisfies \eqref{varsigmaCoefficient} in Appendix~\ref{proofOfzeroInnerProductTheorem}, the coefficient ${\xi _{1,2}}$ is equal to zero, i.e., the channel of UE 1 is orthogonal to that of UE 2, and an IUI-free communication can be obtained. It is also observed that the coefficient ${\xi _{1,2}}$ can be reduced by adjusting the sparsity level $\Gamma$, and thus providing an extra DoF for IUI mitigation as compared to the single-antenna UAV.

 \subsection{Arbitrary Number of UEs}\label{subSectionMultiGeneralUENumber}
 Furthermore, for arbitrary number of UEs, an alternating optimization algorithm similar to  Algorithm~\ref{alg1} is proposed, where the receive beamforming, UAV swarm placement positions, and local positions of MAs are alternately optimized in an iterative manner.

 \subsubsection{Optimization of ${\bf V}$ With Given ${\bf Q}$ and ${\bf B}$}
 The optimization of receive beamforming is the same as \eqref{optimalReceiveBeamforming}, which are omitted for brevity. 
 
 \subsubsection{Optimization of ${{\bf{q}}_l}$ With Given ${\bf V}$, $\left\{ {{\bf{q}}_{l'},\forall l' \ne l} \right\}$, and $\bf B$}
 Let ${{{\bf{\tilde h}}}_{i,l}} \in {{\mathbb C}^{\left( {L - 1} \right)M \times 1}}$ denote the resulting channel after removing ${{\bf{h}}_{i,l}}\in {{\mathbb C}^{M \times 1}} = {\alpha _i}{{\bf{a}}_{i,l}}$ from ${{\bf{h}}_i}$, ${{{\bf{\tilde v}}}_{k,l}} \in {{\mathbb C}^{\left( {L - 1} \right)M \times 1}}$ denote the resulting beamforming after removing ${{\bf{v}}_{k,l}} \in {\mathbb C}^{M \times 1}$ from ${{\bf{v}}_k}$, where ${{\bf{h}}_{i,l}}$ and ${{\bf{v}}_{k,l}}$ denote the $l$-th block of ${{\bf{h}}_i}$ and ${{\bf{v}}_k}$, respectively. For the case of multi-antenna UAV, after some manipulations, ${f_{k,i}}= {\left| {{\bf{v}}_k^H{{\bf{h}}_i}} \right|^2}$ can be expressed as 
 \begin{equation}\label{fkiExpressionMultiAntenna}
 \begin{aligned}
 {f_{k,i}} & = \underbrace {\sum\limits_{l' = 1,l' \ne l}^L {2{{\left| {{\alpha _i}} \right|}^2}{\mathop{\rm Re}\nolimits} \left\{ {{e^{{\rm{j}}\frac{{2\pi }}{\lambda }\left( {{r_{i,l}} - {r_{i,l'}}} \right)}}{\bf{d}}_{i,l}^H{{\bf{V}}_{k,l,l'}}{{\bf{d}}_{i,l'}}} \right\}} }_{{{\tilde g}_{k,i,l}}} \\
 & +\underbrace {{{\left| {{\alpha _i}} \right|}^2}{\bf{d}}_{i,l}^H{{\bf{V}}_{k,l,l}}{{\bf{d}}_{i,l}} + {\bf{\tilde h}}_{i,l}^H{{{\bf{\tilde V}}}_{k,l,l}}{{{\bf{\tilde h}}}_{i,l}}}_{{{\hat g}_{k,i,l}}},
 \end{aligned}
 \end{equation}
 where ${{\bf{d}}_{i,l}} = {[{e^{ - {\rm{j}}\frac{{2\pi }}{\lambda }{\bm{\kappa }}_{i,l}^T{{\bf{b}}_{l,1}}}}, \cdots ,{e^{ - {\rm{j}}\frac{{2\pi }}{\lambda }{\bm{\kappa }}_{i,l}^T{{\bf{b}}_{l,M}}}}]^T}$, ${{\bf{V}}_{k,l,l'}} \in {{\mathbb C}^{M \times M}} \triangleq {{\bf{v}}_{k,l}}{\bf{v}}_{k,l'}^H$, and ${{{\bf{\tilde V}}}_{k,l,l}} \in {{\mathbb C}^{\left( {L - 1} \right)M \times \left( {L - 1} \right)M}} \triangleq {{{\bf{\tilde v}}}_{k,l}}{\bf{\tilde v}}_{k,l}^H$. Moreover, we adopt the quasi-static local AoA approximation, ${{{\hat g}_{k,i,l}}}$ is a constant term, and ${{{\tilde g}_{k,i,l}}}$ can be expressed in terms of ${{{\bf{q}}_l}}$ as 
 \begin{equation}\label{expressiontildegkil}
 \begin{aligned}
 {{\tilde g}_{k,i,l}} &= \sum\limits_{l' = 1,l' \ne l}^L {2{{\left| {{\alpha _i}} \right|}^2}\left| {{\bf{d}}_{i,l}^H{{\bf{V}}_{k,l,l'}}{{\bf{d}}_{i,l'}}} \right| \times } \\
 &\ \ {\cos \left( {\frac{{2\pi }}{\lambda }\left( {{r_{i,l}} - {r_{i,l'}}} \right) + \angle {\bf{d}}_{i,l}^H{{\bf{V}}_{k,l,l'}}{{\bf{d}}_{i,l'}}} \right)}. 
 \end{aligned}
 \end{equation}

 A closer look at \eqref{expressiontildegkil} shows that it has a similar form to \eqref{expressiongkim}, and its lower and upper bounds can be correspondingly obtained. As a result, the optimization of $ {\bf{q}}_l $ for the multi-antenna UAV can follow the similar procedure as in \eqref{subProblemTrajectory3}.

 \subsubsection{Optimization of ${b_{l,m}}$ With Given ${\bf V}$, ${\bf Q}$, and  $\left\{ {{b_{l',m'}},\forall l' \ne l,m' \ne m} \right\}$}
 The sub-problem of (P1) for optimizing $ {b_{l,m}} $ is 
 \begin{equation}\label{subProblemMAPositionMultiAntenna}
 \begin{aligned}
 \mathop {\max }\limits_{ {b_{l,m}},{\rho} }&\ \rho\\
 {\rm{s.t.}}&\ {R_k} \ge \rho,\ \forall k,\\
 &\ {b_{l,m}} \in {\tilde {\cal C}},\\
 &\ {\left| {{b_{l,m}} - {b_{l,m'}}} \right|^2} \ge \tilde d_{\min }^2,\ \forall m' \ne m.
 \end{aligned}
 \end{equation}
 
 By following the similar procedure as in Section~\ref{subsubsectionUAVTrajectory}, problem \eqref{subProblemMAPositionMultiAntenna} can be transformed to 
 \begin{equation}\label{subProblemMAPositionMultiAntenna2}
 \begin{aligned}
 &\mathop {\max }\limits_{{b_{l,m}},\rho,\left\{ {{\eta _k},{\mu _k}} \right\}_{k=1}^K} \ \ \rho \\
 {\rm{s.t.}}&\  {{{\eta _k} - {\mu _k}} }  \ge \rho \ln 2,\ \forall k \in {\cal K},\\
 & {\sigma ^2} + \sum\limits_{i = 1}^K {{P_i}{{\left| {{\bf{v}}_k^H{{\bf{h}}_i}} \right|}^2}}  \ge {\sigma ^2}{e^{{\eta _k}}},\ \forall k \in {\cal K},\\
 & {\sigma ^2} + \sum\limits_{i = 1,i \ne k}^K {{P_i}{{\left| {{\bf{v}}_k^H{{\bf{h}}_i}} \right|}^2}}  \le {\sigma ^2}{e^{{\mu _k}}},\ \forall k \in {\cal K},\\
 &{b_{l,m}} \in {\tilde {\cal C}},\\
 &{\left| {{b_{l,m}} - {b_{l,m'}}} \right|^2} \ge \tilde d_{\min }^2,\ \forall m' \ne m.
 \end{aligned}
 \end{equation}
 
 Let ${{\bf{\tilde h}}_{i,l,m}} \in {{\mathbb C}^{\left( {LM - 1} \right) \times 1}}$ and ${{\bf{\tilde v}}_{k,l,m}} \in {{\mathbb C}^{\left( {LM - 1} \right) \times 1}}$ denote the resulting channel and beamforming vector after removing ${h_{i,l,m}}$ and ${v_{i,l,m}}$, respectively, where ${h_{i,l,m}}$ and ${v_{k,l,m}}$ are the $\left( {\left( {l - 1} \right)M + m } \right)$-th elements of ${{{\bf{h}}_i}}$ and ${{{\bf{v}}_k}}$, respectively. The term ${f_{k,i}} = {\left| {{\bf{v}}_k^H{{\bf{h}}_i}} \right|^2}$ can be further expressed as 
 \begin{equation}
 \begin{aligned}
 {f_{k,i}} = \underbrace {2\left| {{c_{k,i,l,m}}} \right|{\mathop{\rm Re}\nolimits} \left\{ {{e^{ - {\rm{j}}\left( {\frac{{2\pi }}{\lambda }{\Phi _{i,l}}{b_{l,m}} - \angle {c_{k,i,l,m}}} \right)}}} \right\}}_{{g_{k,i,l,m}}} +\\
 \underbrace {  {{\left| {{\alpha _i}} \right|}^2}{{\left| {{v_{k,l,m}}} \right|}^2} + {\bf{\tilde h}}_{i,l,m}^H{{{\bf{\tilde v}}}_{k,l,m}}{\bf{\tilde v}}_{k,l,m}^H{{{\bf{\tilde h}}}_{i,l,m}}}_{{{\bar g}_{k,i,l,m}}},
 \end{aligned}
 \end{equation}
 where ${c_{k,i,l,m}} \triangleq {\bf{\tilde h}}_{i,l,m}^H{{{\bf{\tilde v}}}_{k,l,m}}v_{k,l,m}^ * {\alpha _i}{e^{ - {\rm{j}}\frac{{2\pi }}{\lambda }{r_{i,l}}}}$, and ${{{\bar g}_{k,i,l,m}}}$ is independent of ${{b_{l,m}}}$. Thus, problem \eqref{subProblemMAPositionMultiAntenna2} can be solved similar to \eqref{subProblemTrajectory2}. 
 
 The main procedures for solving problem (P1) are summarized in Algorithm~\ref{alg2}. The complexity for obtaining the receive beamforming in step 3 is ${\cal O}( {K{(LM)^3}} )$. From step 4 to step 6, the complexity is approximately ${\cal O}( {I_1}L{{( {2K})}^3} )$, where $I_1$ denotes the maximum number of iterations required by SCA for convergence. The complexity from step 7 to step 11 is approximately given by ${\cal O}( {I_2}L(M-1){{( {2K})}^3} )$, where $I_2$ denotes the maximum number of iterations to converge required by step 9. As a result, the total computational complexity of Algorithm~\ref{alg2} is ${\cal O}( {{I_3}K{(LM)^3} + {I_1}{I_3}L{{( {2K})}^3}  + {I_2}{I_3}L(M-1){( {2K} )}^3} )$, where $I_3$ denotes the number of iterations required by the alternating optimization for convergence. 
 
 \begin{algorithm}[t]
 \caption{Proposed Alternating Optimization for Solving Problem (P1)}
 \label{alg2}
 \begin{algorithmic}[1]
 \STATE Initialize ${{\bf{Q}}^{\left(0 \right)}}$, ${{\bf{B}}^{\left( 0 \right)}}$, and $\{ {{\bf{v}}_k^{\left( 0 \right)}} \}$ randomly, and let $j=0$.
 \REPEAT
 \STATE For given ${{\bf{Q}}^{\left( j \right)}}$ and ${{\bf{B}}^{\left( j \right)}}$, obtain the optimal beamforming $\{ {{\bf{v}}_k^{\left( {j + 1} \right)}} \}$. 
 \STATE \textbf{for} $l = 1:L$ \textbf{do}
 \STATE \quad Obtain $\{ {\bf{q}}_l^{\left( {j + 1} \right)}\}$ similar to problem \eqref{subProblemTrajectory3}, 
 given \\
 \quad $\{ {\bf{q}}_1^{\left( {j + 1} \right)}, \cdots ,{\bf{q}}_{l - 1}^{\left( {j + 1} \right)},{\bf{q}}_{l}^{\left( j \right)}, \cdots ,{\bf{q}}_L^{\left( j \right)}\}$, ${{\bf{B}}^{\left( j \right)}}$, \\
 \quad  and $\{ {{\bf{v}}_k^{\left( {j + 1} \right)}} \}$.
 \STATE \textbf{end for}
 \STATE \textbf{for} $l = 1:L$ \textbf{do}
 \STATE \quad \textbf{for} $m = 2:M$ \textbf{do}
 \STATE \quad \quad Obtain $ {b_{l,m}^{\left( {j + 1} \right)}} $ given ${{\bf{Q}}^{\left( {j + 1} \right)}}$, $\{ b_{1,2}^{\left( {j + 1} \right)}, \cdots ,$\\
 \quad \quad $b_{l,m - 1}^{\left( {j + 1} \right)},b_{l,m}^{\left( j \right)}, \cdots b_{L,M}^{\left( j \right)}\}$, and $\{ {{\bf{v}}_k^{\left( {j + 1} \right)}} \}$.\\
 \STATE \quad \textbf{end for}
 \STATE \textbf{end for}
 \STATE Update $j=j+1$.
 \UNTIL the fractional increase in the objective function value is below a given threshold $ \epsilon > 0$.
 \end{algorithmic}
 \end{algorithm}
 
 Last, we consider the practical case in the presence of synchronization and position errors. Specifically, let ${\tau _l}$ denote the synchronization error between the local clock of UAV $l$ and the reference clock, which follows a Gaussian distribution \cite{quitin2016scalable,li2024multiobj}. With slight abuse of notations, let ${{\bf{\bar q}}_l}$ denote the nominal position of UAV $l$, and ${\bf{\bar q}}_l^{{\rm{true}}} = {{\bf{\bar q}}_l} + \Delta {{\bf{q}}_l}$ denote its true position, with $\Delta {{\bf{q}}_l} \sim {\cal N}({\bf{0}},\sigma _p^2{\bf{I}})$ being the position error of UAV $l$. By taking into account both the synchronization and position errors, the receive array response vector of UAV for UE $k$ is given by 
 \begin{equation}\label{arrayResponseVectorSingleUAVError}
 {\bf{a}}_{k,l}^{\rm{e}} = {e^{ - {\rm{j}}\frac{{2\pi }}{\lambda }r_{k,l}^{{\rm{true}}}}}{\left[ {{e^{ - {\rm{j}}\frac{{2\pi }}{\lambda }{\varphi _k}\left( {{\bf{\bar q}}_{l,1}^{{\rm{true}}}} \right)}}, \cdots ,{e^{ - {\rm{j}}\frac{{2\pi }}{\lambda }{\varphi _k}\left( {{\bf{\bar q}}_{l,M}^{{\rm{true}}}} \right)}}} \right]^T}\varepsilon _l^{{\rm{syn}}},
 \end{equation}
 where $\varepsilon _l^{{\rm{syn}}} = {e^{ - {\rm{j2}}\pi {f_c}{\tau _l}}}$ denotes the phase shift induced by the synchronization error, with ${f_c}$ being the carrier frequency, and $r_{k,l}^{{\rm{true}}} = \left\| {{\bf{\bar q}}_l^{{\rm{true}}} - {{\bf{w}}_k}} \right\|$. Since the position error is in general much smaller than the link distance, and with the second-order Taylor approximation to $r_{k,l}^{{\rm{true}}}$, the receive array response vector in \eqref{arrayResponseVectorSingleUAVError} is approximated as
 \begin{equation}
 {\bf{a}}_{k,l}^{\rm{e}} \approx {e^{ - {\rm{j}}\frac{{2\pi }}{\lambda }r_{k,l}^{{\rm{second}}}}}{\left[ {{e^{ - {\rm{j}}\frac{{2\pi }}{\lambda }{\varphi _k}\left( {{{{\bf{\bar q}}}_{l,1}}} \right)}}, \cdots ,{e^{ - {\rm{j}}\frac{{2\pi }}{\lambda }{\varphi _k}\left( {{{{\bf{\bar q}}}_{l,M}}} \right)}}} \right]^T}\varepsilon _l^{{\rm{syn}}}\varepsilon _{k,l}^{{\rm{pos}}},
 \end{equation}
 where $\varepsilon _{k,l}^{{\rm{pos}}} = {e^{ - {\rm{j}}\frac{{2\pi }}{\lambda }{\bm{\kappa }}_{k,l}^T\Delta {{\bf{q}}_l}}}$ denotes the phase shift induced by the position error of UAV $l$ for UE $k$. Thus, the channel from UE $k$ to the UAV swarm enabled MA system with synchronization and position errors is 
 \begin{equation}
 {\bf{h}}_k^{\rm{e}} = {\alpha _k}{\left[ {{{\left( {{\bf{a}}_{k,1}^{\rm{e}}} \right)}^T}, \cdots ,{{\left( {{\bf{a}}_{k,L}^{\rm{e}}} \right)}^T}} \right]^T}.
 \end{equation}
 
 Furthermore, we propose a channel estimation-based error compensation scheme to effectively mitigate the adverse effects of synchronization and position errors. It is worth mentioning that both the synchronization and position errors can be regarded as the non-ideal factors of channels from UEs to the UAV swarm enabled MA system. To mitigate the practical issues, when the UAV swarm arrives at the optimized placement positions based on the ideal channel, channel estimation is performed. Specifically, let ${\bf{S}} \in {{\mathbb{C}}^{{T_p} \times K}} = [{{\bf{s}}_1}, \cdots ,{{\bf{s}}_K}]$, where ${{\bf{s}}_k} \in {{\mathbb{C}}^{{T_p} \times 1}}$ denotes the pilot sequence of length $T_p$ for UE $k$, with ${{\bf{S}}^H}{\bf{S}} = {T_p}{{\bf{I}}_K}$. The received signal of the UAV swarm enabled MA system can be expressed as 
 \begin{equation}
 {\bf{Y}} = \sqrt {{p_{{\rm{tr}}}}} {\bf{H}}{{\bf{S}}^T} + {\bf{Z}},
 \end{equation}
 where ${p_{{\rm{tr}}}}$ denotes the training power, ${\bf{H}} \in {\mathbb{C}^{LM \times K}} = \left[{\bf{h}}_1^{\rm{e}}, \cdots ,{\bf{h}}_K^{\rm{e}}\right]$, ${\bf{Z}} \in {{\mathbb{C}}^{LM \times {T_p}}}$ denotes the AWGN matrix. By applying the least-square (LS) estimation, the estimated channel of UE $k$ can be expressed as
 \begin{equation}\label{estimatedChannel}
 {{\bf{\hat h}}_k} = \frac{1}{{\sqrt {{p_{{\rm{tr}}}}} {T_p}}}{\bf{Ys}}_k^*.
 \end{equation}
 
 Based on the estimated channels of UEs, the receive beamforming for UE $k$ is updated to ${{\bf{\hat v}}_k} = \frac{{{\bf{\hat C}}_k^{ - 1}{{{\bf{\hat h}}}_k}}}{{\left\| {{\bf{\hat C}}_k^{ - 1}{{{\bf{\hat h}}}_k}} \right\|}}$, $\forall k$, where ${{\bf{\hat C}}_k} \triangleq {{\bf{I}}_{LM}} + \sum\nolimits_{i = 1,i \ne k}^K {{{\bar P}_i}{{{\bf{\hat h}}}_i}} {\bf{\hat h}}_i^H$. Then, the achievable rate of the channel estimation-based error compensation scheme for UE $k$ is given by 
 \begin{equation}\label{achievableRateError}
 {\hat R_k} = {\log _2}\left( {1 + {{\hat \gamma }_k}} \right) = {\log _2}\left( {1 + \frac{{{{\bar P}_k}{\bf{\hat v}}_k^H{\bf{h}}_k^{\rm{e}}{{\left( {{\bf{h}}_k^{\rm{e}}} \right)}^H}{{{\bf{\hat v}}}_k}}}{{{\bf{\hat v}}_k^H{\bf{C}}_k^{\rm{e}}{{{\bf{\hat v}}}_k}}}} \right),
 \end{equation}
 where ${\bf{C}}_k^{\rm{e}} \triangleq {{\bf{I}}_{LM}} + \sum\nolimits_{i = 1,i \ne k}^K {{{\bar P}_i}{\bf{h}}_i^{\rm{e}}{{({\bf{h}}_i^{\rm{e}})}^H}}$. Accordingly, the minimum achievable rate among $K$ UEs can be obtained.

\section{Numerical Results}\label{sectionNumericalResult}
 In this section, numerical results are provided to verify the performance of the proposed low-altitude UAV swarm enabled MA system. The channel power at the reference distance of ${d_0} = 1$ m is ${\beta _0} =  - 61.4$ dB, and the noise power is ${\sigma ^2} =  - 94$ dBm. The minimum distance to avoid the collision among UAVs is ${d_{\min}} = 1$ m. Moreover, the minimum distance to avoid mutual coupling between adjacent MA elements is ${{\tilde d}_{\min }} = \lambda/2$. The (local) movable region of MAs at each UAV is $\left[ {0,D } \right]$, with $D = 20\lambda$. Unless otherwise stated, $K =3$ UEs are uniformly distributed in a circular area with the center and radius being $\left[{0,0,0}\right]^T$ m and ${R_c} = 600$ m, respectively. The transmit power of each UE is $P_k = 20$ dBm. Unless otherwise stated, the movable region of UAVs is specified by ${{\bar x}_l} \in \left[ {{x_{\min }},{x_{\max }}} \right]$ m, ${{\bar y}_l} \in \left[ {{y_{\min }},{y_{\max }}} \right]$ m, and ${{\bar z}_l} \in \left[ { H, H + z_{\max}} \right]$ m, with ${x_{\min }} = {y_{\min }} =  - 50$ m, ${x_{\max }} = {y_{\max }} = 50$ m, ${z_{\max }} = 40$ m, and $H = 500$ m.

 \subsection{Single-Antenna UAV}
  \begin{figure}[!t]
 \centering
 \centerline{\includegraphics[width=3.5in,height=2.625in]{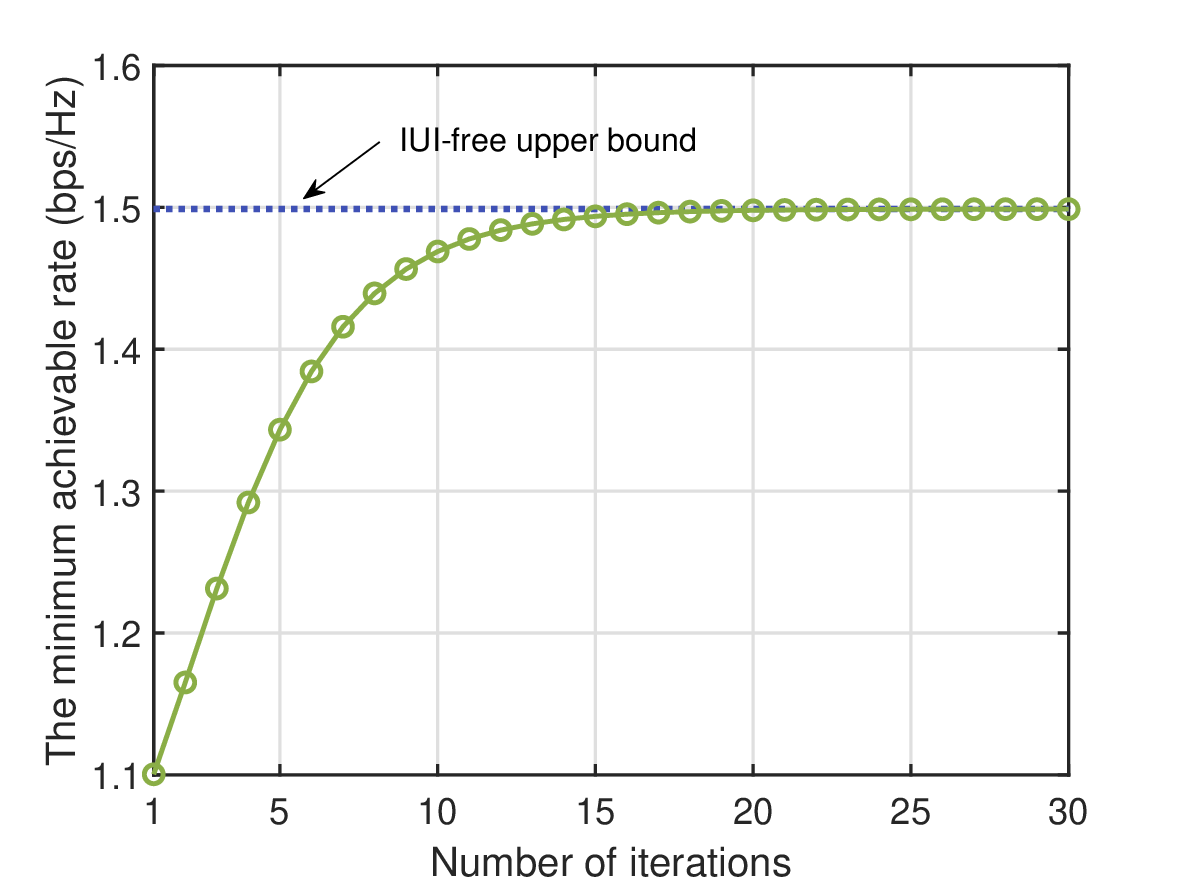}}
 \caption{Convergence behaviour of Algorithm~\ref{alg1}.}
 \label{fig:convergencePerformanceSingleAntenna}
 \end{figure}
 
 First, we consider the case of single-antenna UAVs, with $L = 4$. Fig.~\ref{fig:convergencePerformanceSingleAntenna} shows the convergence behaviour of Algorithm~\ref{alg1}. The upper bound of IUI-free communication is also provided for comparison, i.e., the result given in \eqref{minrateTwoUESingleAntenna}. It is observed that Algorithm~\ref{alg1} yields a non-decreasing minimum achievable rate, and finally approaches the converged solution that is close to the IUI-free upper bound. 
 
 \begin{figure}[!t]
 \centering
 \centerline{\includegraphics[width=3.5in,height=2.625in]{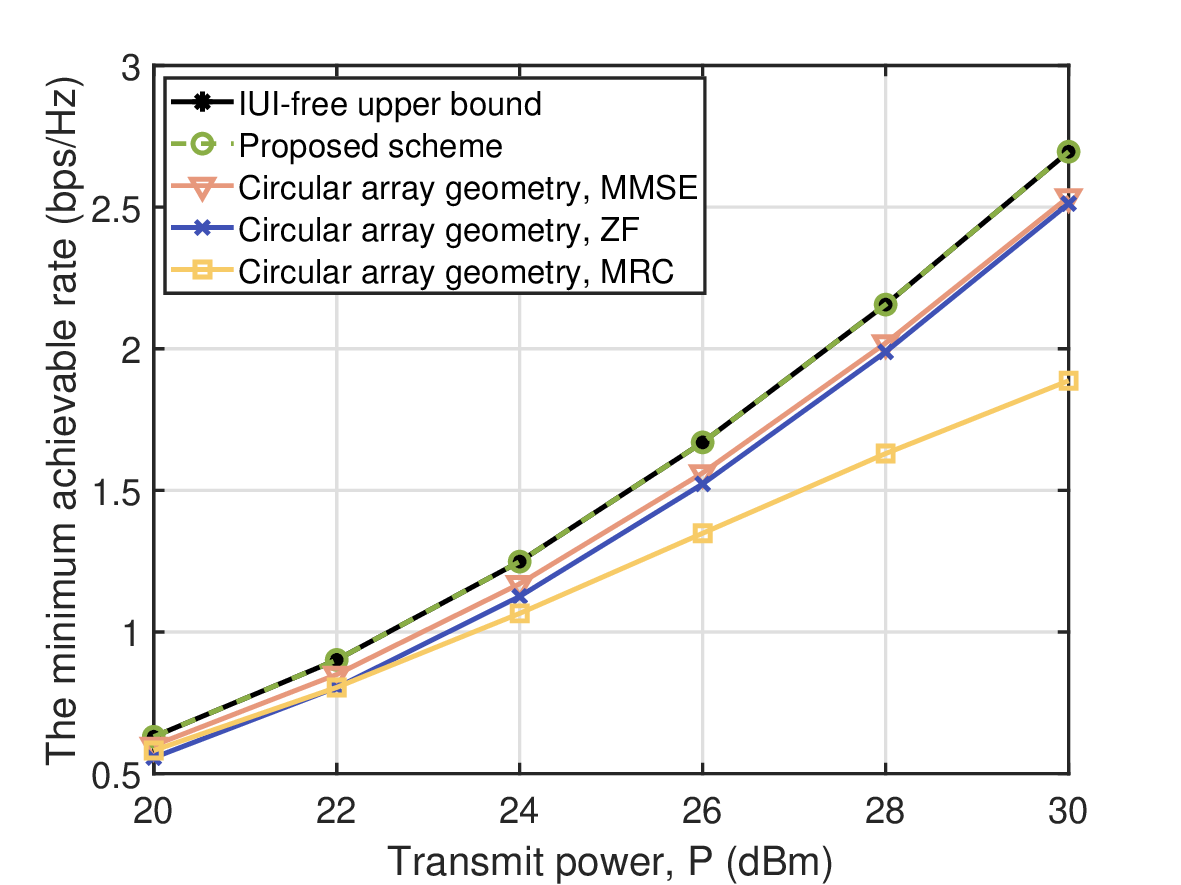}}
 \caption{The minimum achievable rate versus the transmit power of each UE for the two-UE communication in the single-antenna UAV case.}
 \label{fig:achievableRateVersusTranspitPowerTwoUESingle}
 \end{figure}
 
\begin{figure}[!t]
 \centering
 \centerline{\includegraphics[width=3.5in,height=2.625in]{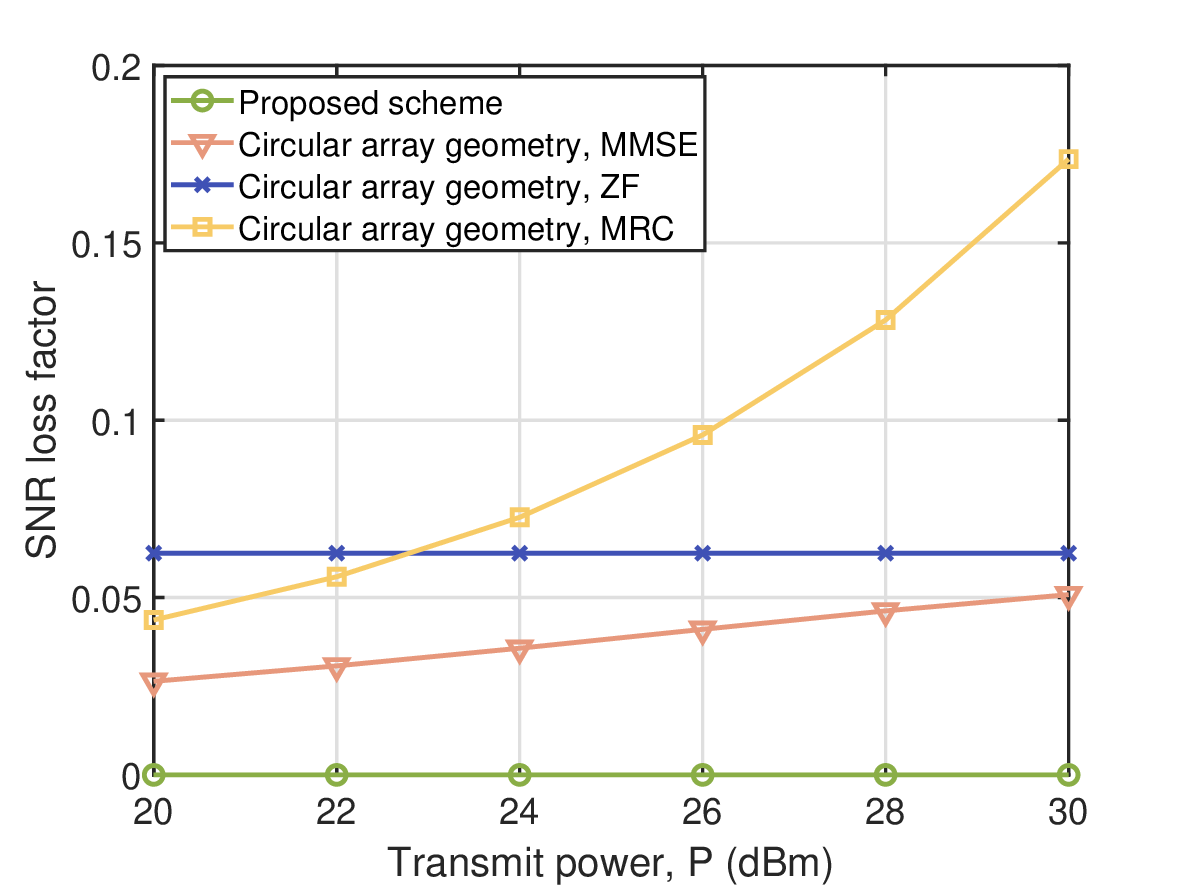}}
 \caption{The SNR loss factor versus the transmit power for the two-UE communication in the single-antenna UAV case.}
 \label{fig:SNRLossFactorVersusTransmitPower}
 \end{figure}

 Fig.~\ref{fig:achievableRateVersusTranspitPowerTwoUESingle} shows the minimum achievable rate versus the transmit power of each UE for the two-UE communication, by considering the UE directions $\left( {{\theta _1},{\phi _1}} \right) = \left( {{{30}^ \circ },{{60}^ \circ }} \right)$ and $\left( {{\theta _2},{\phi _2}} \right) = \left( {{{0}^ \circ },{{0}^ \circ }} \right)$, respectively. The movable region of UAVs is specified by ${{\bar x}_l} \in \left[ {{x_{\min }},{x_{\max }}} \right]$ m, ${{\bar y}_l} \in \left[ {{y_{\min }},{y_{\max }}} \right]$ m, and ${{\bar z}_l} \in \left[ { H, H + z_{\max}} \right]$ m, with ${x_{\min }} = {y_{\min }} =  - 20$ m, ${x_{\max }} = {y_{\max }} = 20$ m, ${z_{\max }} = 10$ m, and $H = 1000$ m. For comparison, the benchmark scheme of circular array geometry is considered, i.e., the UAV swarm cooperatively forms a circular array, with their placement positions given by 
 \begin{equation}\label{circularTrajectory}
 {{\bf{\bar q}}_l}{\rm =} {{\bf{q}}_r} + \left[ R\cos \left( {\frac{{2\pi }}{L}\left( {l - 1} \right)} \right), {R\sin \left( {\frac{{2\pi }}{L}\left( {l - 1} \right)} \right),0} \right]^T,\ \forall l,
 \end{equation}
 where $R$ denotes the radius of circular array geometry and is set as $R = \left( {{x_{\max }} - {x_{\min }}} \right)/2 = \left( {{y_{\max }} - {y_{\min }}} \right)/2$. With the circular array geometry, the MMSE, ZF and MRC beamforming schemes are respectively considered. It is observed that the proposed UAV swarm enabled MA system achieves an IUI-free communication, thanks to the flexible placement position optimization. Besides, the proposed scheme outperforms the benchmark schemes of circular array geometry with MMSE, ZF, and MRC beamforming. This is expected since the proposed scheme is able to completely orthogonalize the channels of two UEs, while achieving the full beamforming gain at each UE. Specifically, Fig.~\ref{fig:SNRLossFactorVersusTransmitPower} shows the SNR loss factor versus the transmit power for UE 1. It is observed that the proposed scheme always enjoys a zero SNR loss factor. By contrast, the circular array geometry with MMSE and MRC beamforming schemes experience increased SNR loss factors as the transmit power increases, as a result of suffering from more severe IUI, especially for the MRC beamforming scheme.

 \begin{figure}[!t]
 \centering
 \centerline{\includegraphics[width=3.5in,height=2.625in]{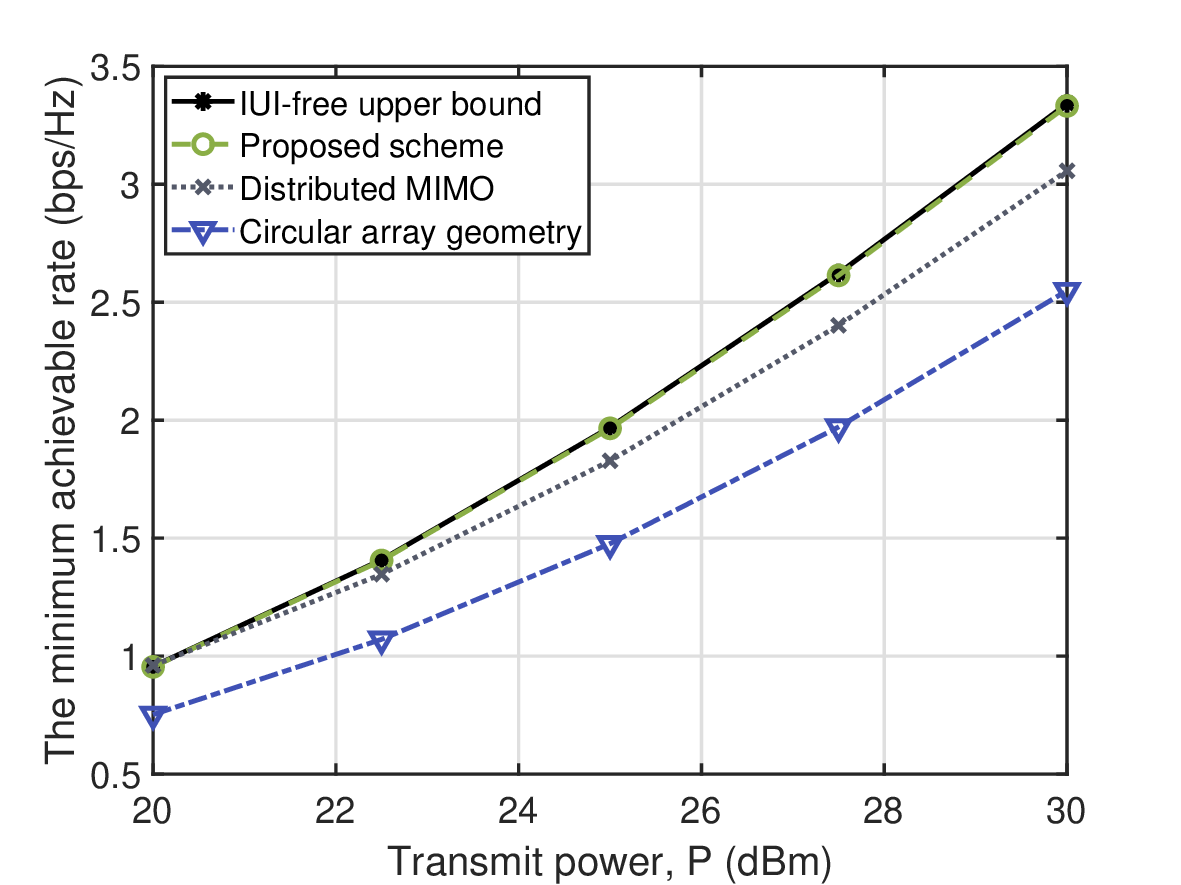}}
 \caption{The minimum achievable rate versus the transmit power of each UE for three UEs in the single-antenna UAV case.}
 \label{fig:achievableRateVersusTransmitPowerSingleDistributed}
 \end{figure}
 
 For further comparison, Fig.~\ref{fig:achievableRateVersusTransmitPowerSingleDistributed} shows the minimum achievable rate versus the transmit power of each UE for $K =3$ UEs. Since MMSE beamforming achieves the balance between reducing the interference and noise enhancement, the benchmark scheme of circular array geometry with MMSE beamforming is considered in the following. Moreover, the benchmark scheme of distributed MIMO is considered \cite{ammar2022user}, where four single-antenna access points (APs) cooperatively serve the UEs, and their locations are given by ${[{R_c}/2,{R_c}/2,0]^T}$, ${[ - {R_c}/2,{R_c}/2,0]^T}$, ${[ - {R_c}/2,-{R_c}/2,0]^T}$, ${[{R_c}/2, - {R_c}/2,0]^T}$, respectively, with ${R_c} = 1000$ m. It is firstly observed that the minimum achievable rate of both the proposed and benchmark schemes increase as the transmit power increases, as expected. Besides, similar to the case of two UEs, the proposed UAV swarm enabled MA system yields a comparable performance to the IUI-free upper bound, and is significantly superior to the benchmark scheme of circular array geometry with MMSE beamforming. This is expected since the proposed scheme can strike a good balance between the beamforming gain improvement and IUI mitigation via the flexible UAV placement position adjustment. It is also observed that the proposed scheme outperforms the benchmark scheme of distributed MIMO as the transmit power increases. This is expected since although the distributed MIMO deploys the geographically separated APs, their fixed-position antennas cannot exploit the channel variation in the spatial domain for mitigating the IUI.
 
 \begin{figure}[!t]
 \centering
 \centerline{\includegraphics[width=3.5in,height=2.625in]{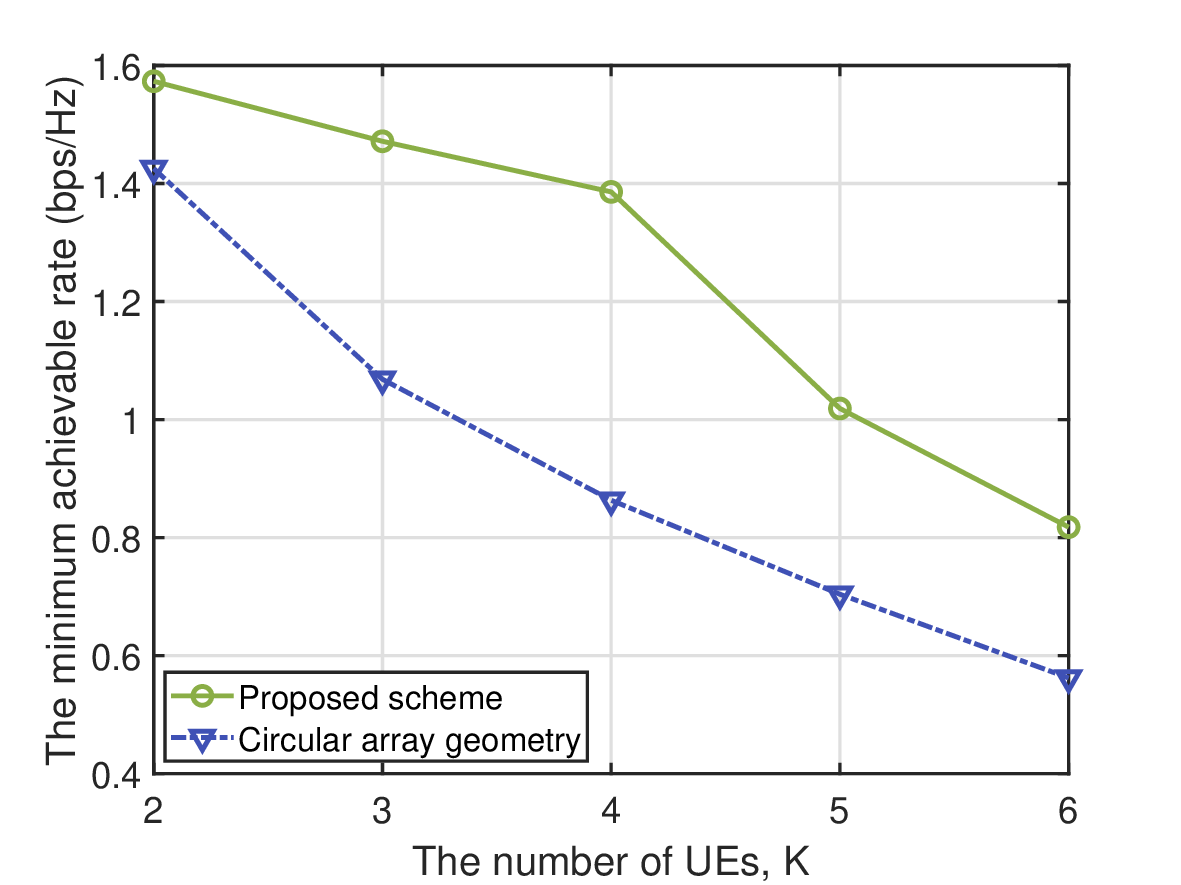}}
 \caption{The minimum achievable rate versus the number of UEs in the single-antenna UAV case.}
 \label{fig:achievableRateVersusUENumberSingle}
 \end{figure}
 
 Fig.~\ref{fig:achievableRateVersusUENumberSingle} shows the minimum achievable rate versus the number of UEs, $K$. It is observed that the performance of both schemes decrease as the number of UEs increases. This is because more UEs will cause a severer IUI issue, especially when the number of UAVs/antennas is smaller than that of UEs, e.g., $K > 4$. It is also observed that the minimum achievable rate of the proposed UAV swarm enabled MA system always surpasses the benchmark scheme of circular array geometry, thanks to the flexible UAV placement position adjustment to significantly reduce the channel correlation and thus the IUI among UEs.

 \subsection{Multi-Antenna UAV} 

\begin{figure}[!t]
 \centering
 \centerline{\includegraphics[width=3.5in,height=2.625in]{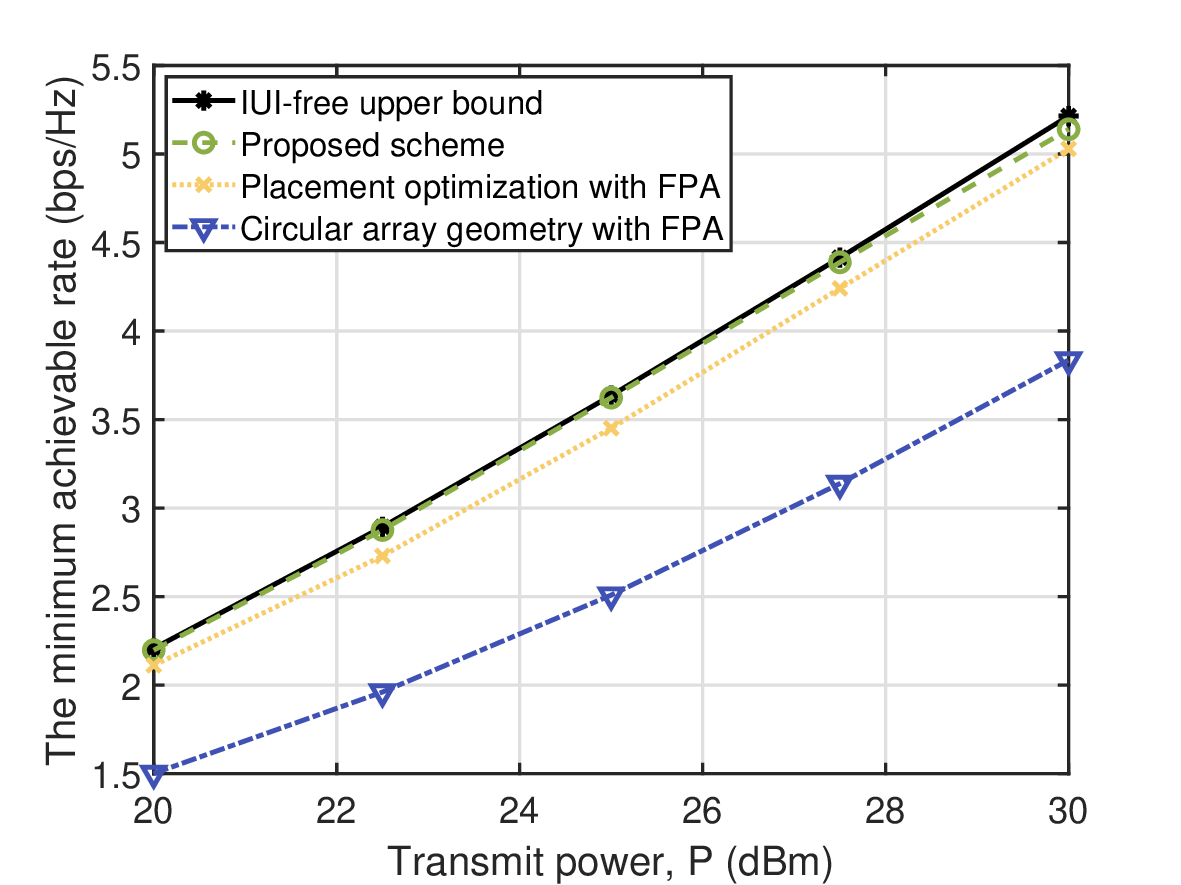}}
 \caption{The minimum achievable rate versus the transmit power of each UE in the multi-antenna UAV case.}
 \label{fig:achievableRateVersusTranspitPowerMultiFPA}
 \end{figure}

 Next, we consider the case of multi-antenna UAVs, where the numbers of UAVs and MAs per UAV are $L=2$ and $M=4$, respectively. For comparison, the following two benchmark schemes are considered: 1) Circular array geometry with FPA: each UAV is equipped with an FPA array (linear) and UAV swarm adopts the circular geometry given in \eqref{circularTrajectory}; 2) Placement optimization with FPA: each UAV is equipped with the FPA array as above and UAV swarm placement is optimized similar to Section~\ref{subSectionMultiGeneralUENumber}. Fig.~\ref{fig:achievableRateVersusTranspitPowerMultiFPA} shows the minimum achievable rate versus the transmit power of each UE for the case with $K =3$ UEs. It is observed that the performance of UAV swarm enabled MA system is very close to that of the IUI-free communication, and its performance gains over the other two benchmark schemes become even more significant as the transmit power increases. This is due to the two-level mobility brought by UAV swarm enabled MA system, i.e., the intrinsic mobility of UAV swarm and additional antenna position adjustment of MA. Specifically, it can be observed that the scheme of placement optimization with FPA yields a better performance than that of circular UAV geometry with FPA, which demonstrates the importance of UAV swarm placement optimization for performance improvement. Moreover, compared to the scheme of placement optimization with FPA, considerable performance gain is achieved for UAV swarm enabled MA system, thanks to the additional antenna position adjustment of MA. The above results verify the advantages of two-level mobility brought by UAV swarm enabled MA system, i.e., it can fully exploit channel variation for balancing the beamforming gain improvement and IUI reduction. 

 \begin{figure}
 \centering
 \subfigure[$K = 2$]{
 \begin{minipage}[t]{0.5\textwidth}
 \centering
 \centerline{\includegraphics[width=3.5in,height=2.625in]{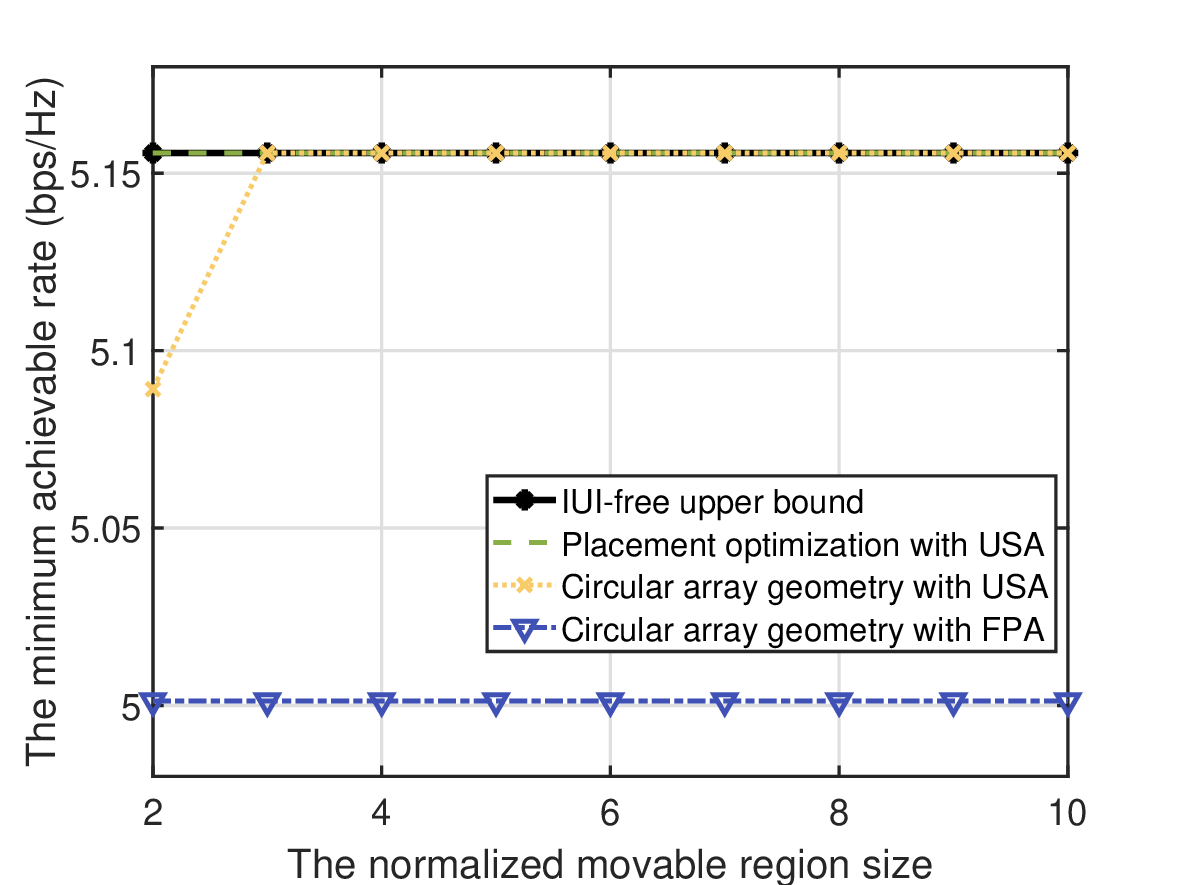}}
 \end{minipage}
 }
 \subfigure[$K = 3$]{
 \begin{minipage}[t]{0.5\textwidth}
 \centering
 \centerline{\includegraphics[width=3.5in,height=2.625in]{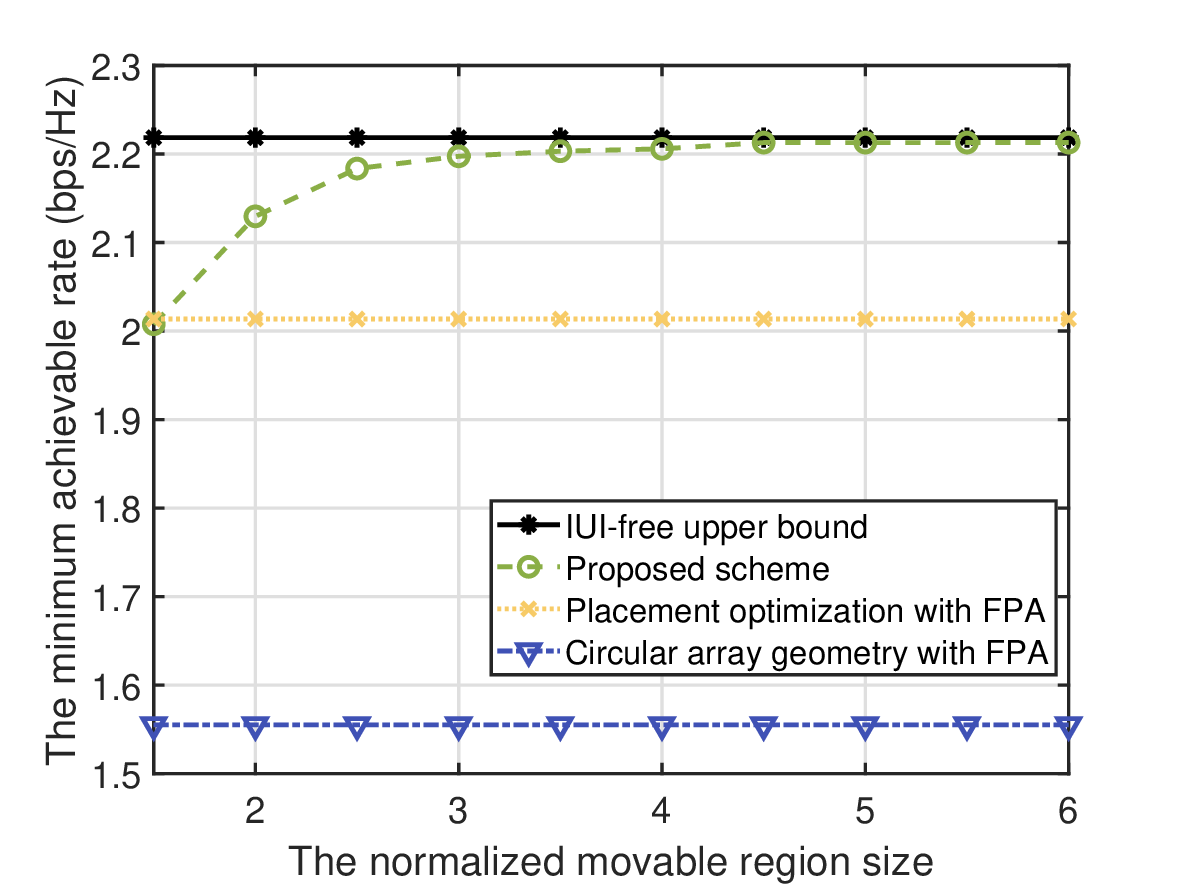}}
 \end{minipage}
 }
 \caption{The minimum achievable rate versus the normalized movable region size in the multi-antenna UAV case.}
 \label{fig:achievableRateVersusMovableRegionSize}
 \end{figure}

 Fig.~\ref{fig:achievableRateVersusMovableRegionSize} studies the impact of the normalized movable region size $D/\lambda$ on the minimum achievable rate. For $K =2$, the transmit power of each UE is ${P_k} = 35$ dBm, the UE directions remain consistent with those in Fig.~\ref{fig:achievableRateVersusTranspitPowerTwoUESingle}, and the following three schemes are considered: 1) Placement optimization with USA: each UAV is equipped with a USA and UAV swarm adopts the placement given in \eqref{zeroInnerProductTrajectory} and \eqref{minimumInteger}; 2) Circular array geometry with USA: each UAV is equipped with a USA and UAV swarm adopts the circular geometry given in \eqref{circularTrajectory}; 3) Circular array geometry with FPA: Similar to 2), but using FPA instead USA. For USA, the sparsity level can be dynamically adjusted. It is firstly observed from Fig.~\ref{fig:achievableRateVersusMovableRegionSize}(a) that the scheme of placement optimization with USA directly achieves the IUI-free communication. This is because the placement given in \eqref{zeroInnerProductTrajectory} and \eqref{minimumInteger} orthogonalizes the channels of two UEs. On the other hand, as the normalized movable region size increases, the scheme of circular array geometry with USA also achieves the IUI-free communication, thanks to the extra DoF of sparsity level adjustment for completely eliminating the IUI. Moreover, for $K = 3$, it is observed from Fig.~\ref{fig:achievableRateVersusMovableRegionSize}(b) that the performance of two FPA schemes remain unchanged as the normalized movable region size increases, as expected. By contrast, the performance of the proposed UAV swarm enabled MA system first increases and ultimately converges as the normalized movable region size increases. This is mainly attributed to the fact that with an enlarged movable region size, MA array has a larger spatial DoF to create favorable channels for performance improvement. However, this does not mean that the performance gain of UAV swarm enabled MA system over other benchmark schemes will continuously increase, since it is upper-bounded by the performance under IUI-free communication.

 \begin{figure}[!t]
 \centering
 \centerline{\includegraphics[width=3.5in,height=2.625in]{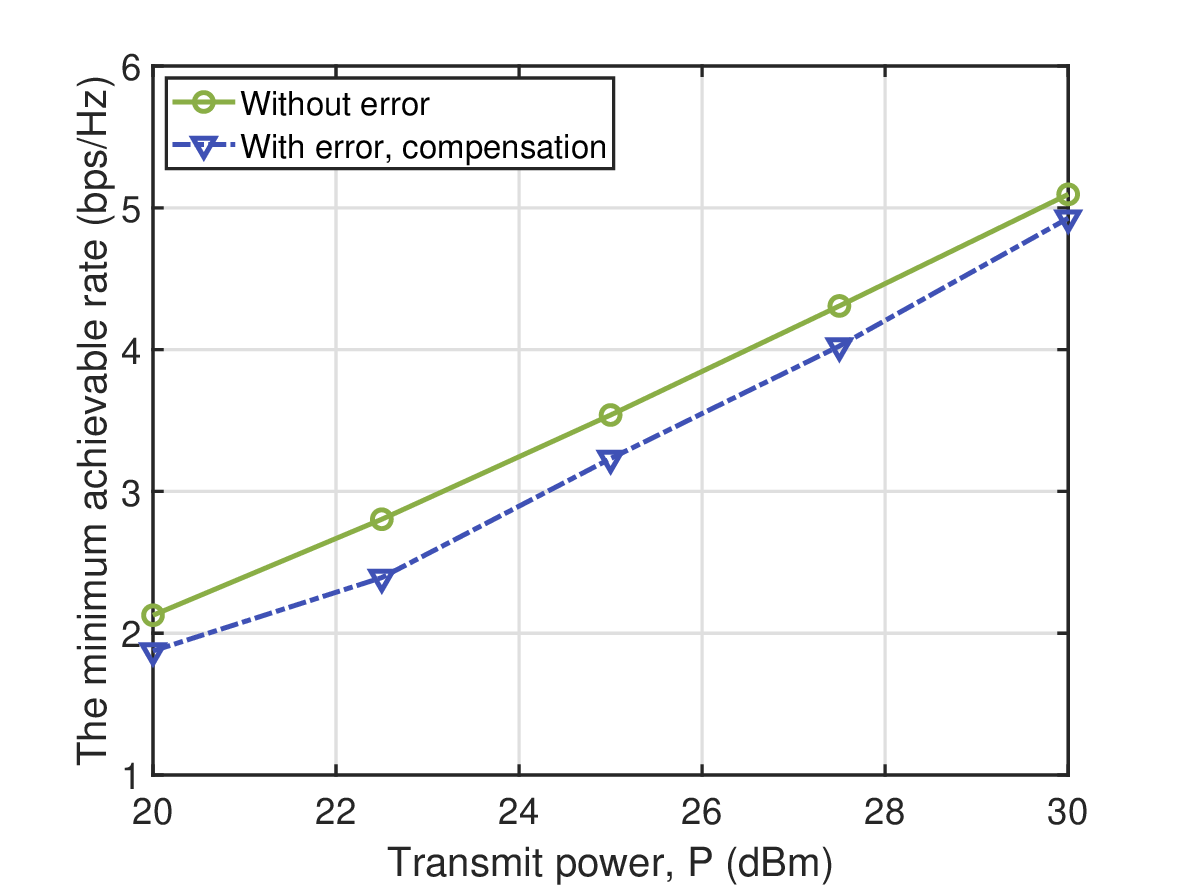}}
 \caption{Comparison of the minimum achievable rate for the cases with and without synchronization and position errors.}
 \label{fig:achievableRateVersusTranspitPowerMultiFPAError}
 \end{figure}
 Last, Fig.~\ref{fig:achievableRateVersusTranspitPowerMultiFPAError} compares the minimum achievable rate for the cases with and without synchronization and position errors, where for the former, the proposed channel estimation-based error compensation scheme is considered. The standard deviations of synchronization and position errors are $31.6$ $\mu$s, and ${\sigma _p} = \lambda $, respectively. The length of pilot sequence is ${T_p} = 60$, and the training power is ${p_{{\rm{tr}}}} = P$. It is observed that the proposed channel estimation-based error compensation scheme yields a comparable performance to the case without errors, which demonstrates the effectiveness of the proposed scheme in compensating for synchronization and position uncertainties. 
\section{Conclusion}\label{sectionConclusion}
 This paper proposed a novel UAV swarm enabled two-level MA system to support low-altitude economy, by exploiting the controllable mobility of UAV and antenna position adjustment of MA. An optimization problem was formulated to maximize the minimum achievable rate over all ground UEs, by jointly optimizing 3D UAV swarm placement positions, their individual MAs' positions, and receive beamforming for UEs. To gain useful insights, we first considered the special case of single-antenna UAV. It was shown that the resulting SNR of single UE communication is independent of UAV swarm array geometry, and the optimal UAV swarm placement positions were derived in closed-form for two-UE communication. For arbitrary number of UEs, an alternating optimization algorithm was proposed to efficiently solve the formulated non-convex problem. Moreover, the results of single-antenna UAV were extended to the general case with multi-antenna UAV. Numerical results demonstrated that significant performance gains can be achieved for the proposed UAV swarm enabled MA system over various benchmarks, thanks to the two-level antenna mobility to create more favorable channels. 

 In addition, there are many important research directions that could be pursued in the future. For instance, the channel modeling incorporating the array misalignment and orientation jitter issues, as well as the extension to multi-path channels, are worthwhile to investigate. Besides achievable rate, energy consumption and latency are the important performance metrics for UAV swarm enabled MA system, while designing the energy-efficient and low-latency UAV system needs more in-depth studies. Moreover, the flight control mechanisms of UAV swarm to guarantee the safety and mission execution are important to investigate in the future. Last, how to integrate IRS into UAV swarm enabled MA systems is an interesting direction that deserves further investigation.

\begin{appendices}
\section{Proof of Theorem~\ref{zeroInnerProductTheorem}}\label{proofOfzeroInnerProductTheorem}
 By substituting ${{\bf{q}}_l} ={{\bf{q}}_1} + \left( {l - 1} \right)\times$ $\frac{{\left( {{\varsigma}  + 1/L} \right)\lambda }}{{{{\left\| {{{\bm{\kappa }}_k} - {{\bm{\kappa }}_{k'}}} \right\|}^2}}}  \left( {{{\bm{\kappa }}_k} - {{\bm{\kappa }}_{k'}}} \right)$ into \eqref{correlationCoefficient}, where $\varsigma$ is a common coefficient to all the UAVs, given by 
 \begin{equation}\label{varsigmaCoefficient}
 \varsigma  \in \left\{ {\left. {\frac{\nu }{L}} \right|\nu  \in {\mathbb Z},\bmod \left( {\nu  + 1,L} \right) \ne 0} \right\},
 \end{equation}
 it can be verified that the objective value of \eqref{correlationCoefficient} is equal to zero.  Moreover, the distance between UAV $l$ and $l'$ is given by 
 \begin{equation}
 \begin{aligned}
 &\left\| {{{\bf{q}}_l} - {{\bf{q}}_{l'}}} \right\| = \left\| {\left( {l - l'} \right)\left( {\varsigma  + 1/L} \right)\lambda \frac{{\left( {{{\bm{\kappa }}_k} - {{\bm{\kappa }}_{k'}}} \right)}}{{{{\left\| {{{\bm{\kappa }}_k} - {{\bm{\kappa }}_{k'}}} \right\|}^2}}}} \right\|\\
 &\ge \left\| {\left( {\varsigma  + 1/L} \right)\lambda \frac{{\left( {{{\bm{\kappa }}_k} - {{\bm{\kappa }}_{k'}}} \right)}}{{{{\left\| {{{\bm{\kappa }}_k} - {{\bm{\kappa }}_{k'}}} \right\|}^2}}}} \right\| = \frac{{\lambda \left| { \varsigma  + 1/L}  \right|}}{{\left\| {{{\bm{\kappa }}_k} - {{\bm{\kappa }}_{k'}}} \right\|}}.
 \end{aligned}
 \end{equation}
 To satisfy the minimum safe distance constraint for UAVs, a feasible solution of $\varsigma $ is thus given by \eqref{minimumInteger}. This completes the proof of Theorem~\ref{zeroInnerProductTheorem}.

 \section{Proof of Lemma \ref{quadraticSurrogatelemma}}\label{proofOfquadraticSurrogatelemma}
 With ${g_{k,i,l}}$ given in \eqref{expressiongkim}, its lower and upper bounds can be constructed based on the second-order Taylor expansion \cite{sun2016majorization,ma2024mimo,wang2024throughput}. Specifically, the gradient of ${g_{k,i,l}}$ over ${{{\bf{q}}_l}}$ is $\nabla {g_{k,i,l}} = {\left[ {\frac{{\partial {g_{k,i,l}}}}{{\partial {x_l}}},\frac{{\partial {g_{k,i,l}}}}{{\partial {y_l}}},\frac{{\partial {g_{k,i,l}}}}{{\partial {z_l}}}} \right]^T}$, where 
 \begin{equation}\label{gradientxm}
 \begin{aligned}
 &\frac{{\partial {g_{k,i,l}}}}{{\partial {x_l}}} =  - \frac{{2\pi }}{\lambda }\sum\limits_{l' = 1,l' \ne l}^L {2{{\left| {{\alpha _i}} \right|}^2}\left| {{V_{k,l,l'}}} \right|\frac{{\partial {r_{i,l}}}}{{\partial {x_l}}}} \\
 &\ \ \ \ \ \ \ \ \ \ \times {\rm{sin}}\left( {\frac{{2\pi }}{\lambda }\left( {{r_{i,l}} - {r_{i,l'}}} \right) + \angle {V_{k,l,l'}}} \right),
 \end{aligned}
 \end{equation}
 \begin{equation}\label{gradientym}
 \begin{aligned}
 &\frac{{\partial {g_{k,i,l}}}}{{\partial {y_l}}} =  - \frac{{2\pi }}{\lambda }\sum\limits_{l' = 1,l' \ne l}^L {2{{\left| {{\alpha _i}} \right|}^2}\left| {{V_{k,l,l'}}} \right|\frac{{\partial {r_{i,l}}}}{{\partial {y_l}}}} \\
 &\ \ \ \ \ \ \ \ \ \ \times {\rm{sin}}\left( {\frac{{2\pi }}{\lambda }\left( {{r_{i,l}} - {r_{i,l'}}} \right) + \angle {V_{k,l,l'}}} \right),
 \end{aligned}
 \end{equation}
 \begin{equation}\label{gradientzm}
 \begin{aligned}
 &\frac{{\partial {g_{k,i,l}}}}{{\partial {z_l}}} =  - \frac{{2\pi }}{\lambda }\sum\limits_{l' = 1,l' \ne l}^L {2{{\left| {{\alpha _i}} \right|}^2}\left| {{V_{k,l,l'}}} \right|\frac{{\partial {r_{i,l}}}}{{\partial {z_l}}}} \\
 &\ \ \ \ \ \ \ \ \ \ \times {\rm{sin}}\left( {\frac{{2\pi }}{\lambda }\left( {{r_{i,l}} - {r_{i,l'}}} \right) + \angle {V_{k,l,l'}}} \right),
 \end{aligned}
 \end{equation}
 where $\frac{{\partial {r_{i,l}}}}{{\partial {x_l}}} = {\Phi _i} + \frac{{{x_l} - \left( {{x_l}{\Phi _i} + {y_l}{\Psi _i} + {z_l}{\Theta _i}} \right){\Phi _i}}}{{{r_i}}}$, $\frac{{\partial {r_{i,l}}}}{{\partial {y_l}}} = {\Psi _i} + \frac{{{y_l} - \left( {{x_l}{\Phi _i} + {y_l}{\Psi _i} + {z_l}{\Theta _i}} \right){\Psi _i}}}{{{r_i}}}$, and $\frac{{\partial {r_{i,l}}}}{{\partial {z_l}}} = {\Theta _i} + \frac{{{z_l} - \left( {{x_l}{\Phi _i} + {y_l}{\Psi _i} + {z_l}{\Theta _i}} \right){\Theta _i}}}{{{r_i}}}$, respectively. 
 
 Besides, the Hessian matrix of ${{g_{k,i,l}}}$ over ${{{\bf{q}}_l}}$ is 
 \begin{equation}\label{HessianMatrix}
 {\nabla ^2}{g_{k,i,l}} {\rm = }\left[ {\begin{array}{*{20}{c}}
 {\frac{{{\partial ^2}{g_{k,i,l}}}}{{\partial {x_l}\partial {x_l}}}}&{\frac{{{\partial ^2}{g_{k,i,l}}}}{{\partial {x_l}\partial {y_l}}}}&{\frac{{{\partial ^2}{g_{k,i,l}}}}{{\partial {x_l}\partial {z_l}}}}\\
 {\frac{{{\partial ^2}{g_{k,i,l}}}}{{\partial {y_l}\partial {x_l}}}}&{\frac{{{\partial ^2}{g_{k,i,l}}}}{{\partial {y_l}\partial {y_l}}}}&{\frac{{{\partial ^2}{g_{k,i,l}}}}{{\partial {y_l}\partial {z_l}}}}\\
 {\frac{{{\partial ^2}{g_{k,i,l}}}}{{\partial {z_l}\partial {x_l}}}}&{\frac{{{\partial ^2}{g_{k,i,l}}}}{{\partial {z_l}\partial {y_l}}}}&{\frac{{{\partial ^2}{g_{k,i,l}}}}{{\partial {z_l}\partial {z_l}}}}
 \end{array}} \right],
 \end{equation}
 where 
 \begin{equation}
 \begin{aligned}
 &\frac{{{\partial ^2}{g_{k,i,l}}}}{{\partial {x_l}\partial {x_l}}} =  - \frac{{2\pi }}{\lambda }\sum\limits_{l' = 1,l' \ne l}^L {2{{\left| {{\alpha _i}} \right|}^2}\left| {{V_{k,l,l'}}} \right| \times } \\
 &\ \ \left[ {\frac{{1 - \Phi _i^2}}{{{r_i}}}{\rm{sin}}\left( {\frac{{2\pi }}{\lambda }\left( {{r_{i,l}} - {r_{i,l'}}} \right) + \angle {V_{k,l,l'}}} \right) + } \right.\\
 &\ \ \ \left. {\frac{{2\pi }}{\lambda }{{\left( {\frac{{\partial {r_{i,l}}}}{{\partial {x_l}}}} \right)}^2}\cos \left( {\frac{{2\pi }}{\lambda }\left( {{r_{i,l}} - {r_{i,l'}}} \right) + \angle {V_{k,l,l'}}} \right)} \right],
 \end{aligned}
 \end{equation}
 \begin{equation}
 \begin{aligned}
 &\frac{{{\partial ^2}{g_{k,i,l}}}}{{\partial {x_l}\partial {y_l}}} =  - \frac{{2\pi }}{\lambda }\sum\limits_{l' = 1,l' \ne l}^L {2{{\left| {{\alpha _i}} \right|}^2}\left| {{V_{k,l,l'}}} \right| \times } \\
 &\ \ \left[ {\frac{{ - {\Phi _i}{\Psi _i}}}{{{r_i}}}{\rm{sin}}\left( {\frac{{2\pi }}{\lambda }\left( {{r_{i,l}} - {r_{i,l'}}} \right) + \angle {V_{k,l,l'}}} \right) + } \right.\\
 &\ \ \ \left. {\frac{{2\pi }}{\lambda }\frac{{\partial {r_{i,l}}}}{{\partial {x_l}}}\frac{{\partial {r_{i,l}}}}{{\partial {y_l}}}\cos \left( {\frac{{2\pi }}{\lambda }\left( {{r_{i,l}} - {r_{i,l'}}} \right) + \angle {V_{k,l,l'}}} \right)} \right],
 \end{aligned}
 \end{equation}
 \begin{equation}
 \begin{aligned}
 &\frac{{{\partial ^2}{g_{k,i,l}}}}{{\partial {x_l}\partial {z_l}}} =  - \frac{{2\pi }}{\lambda }\sum\limits_{l' = 1,l' \ne l}^L {2{{\left| {{\alpha _i}} \right|}^2}\left| {{V_{k,l,l'}}} \right| \times } \\
 &\ \ \left[ {\frac{{ - {\Phi _i}{\Theta _i}}}{{{r_i}}}{\rm{sin}}\left( {\frac{{2\pi }}{\lambda }\left( {{r_{i,l}} - {r_{i,l'}}} \right) + \angle {V_{k,l,l'}}} \right) + } \right.\\
 &\ \ \ \left. {\frac{{2\pi }}{\lambda }\frac{{\partial {r_{i,l}}}}{{\partial {x_l}}}\frac{{\partial {r_{i,l}}}}{{\partial {z_l}}}\cos \left( {\frac{{2\pi }}{\lambda }\left( {{r_{i,l}} - {r_{i,l'}}} \right) + \angle {V_{k,l,l'}}} \right)} \right],
 \end{aligned}
 \end{equation}
 and other elements can be similarly obtained based on \eqref{gradientxm}-\eqref{gradientzm}, which are omitted for brevity.

 Moreover, with \eqref{HessianMatrix}, we have \eqref{hessianMatrixBound}, shown at the top of the next page. 
 \newcounter{mytempeqncnt1}
 \begin{figure*}
 \normalsize
 \setcounter{mytempeqncnt1}{\value{equation}}
 \begin{equation}\label{hessianMatrixBound}
 \begin{aligned}
 &\left\| {{\nabla ^2}{g_{k,i,l}}} \right\|_2^2 \le \left\| {{\nabla ^2}{g_{k,i,l}}} \right\|_F^2\; = {\left( {\frac{{{\partial ^2}{g_{k,i,l}}}}{{\partial {x_l}\partial {x_l}}}} \right)^2} +  \cdots  + {\left( {\frac{{{\partial ^2}{g_{k,i,l}}}}{{\partial {z_l}\partial {z_l}}}} \right)^2} \le {\left( {\frac{{2\pi }}{\lambda }\sum\limits_{l' = 1,l' \ne l}^L {2{{\left| {{\alpha _i}} \right|}^2}\left| {{V_{k,l,l'}}} \right|} } \right)^2} \times \\
 &\left[ {{{\left( {\frac{{1 - \Phi _i^2}}{{{r_i}}}} \right)}^2} + {{\left( {\frac{{2\pi }}{\lambda }{{\left( {\frac{{\partial {r_{i,l}}}}{{\partial {x_l}}}} \right)}^2}} \right)}^2} + {{\left( {\frac{{{\Phi _i}{\Psi _i}}}{{{r_i}}}} \right)}^2} + {{\left( {\frac{{2\pi }}{\lambda }\frac{{\partial {r_{i,l}}}}{{\partial {x_l}}}\frac{{\partial {r_{i,l}}}}{{\partial {y_l}}}} \right)}^2} + {{\left( {\frac{{{\Phi _i}{\Theta _i}}}{{{r_i}}}} \right)}^2} + {{\left( {\frac{{2\pi }}{\lambda }\frac{{\partial {r_{i,l}}}}{{\partial {x_l}}}\frac{{\partial {r_{i,l}}}}{{\partial {z_l}}}} \right)}^2} + } \right.\\
 &{\left( {\frac{{{\Psi _i}{\Phi _i}}}{{{r_i}}}} \right)^2} + {\left( {\frac{{2\pi }}{\lambda }\frac{{\partial {r_{i,l}}}}{{\partial {y_l}}}\frac{{\partial {r_{i,l}}}}{{\partial {x_l}}}} \right)^2} + {\left( {\frac{{1 - \Psi _i^2}}{{{r_i}}}} \right)^2} + {\left( {\frac{{2\pi }}{\lambda }{{\left( {\frac{{\partial {r_{i,l}}}}{{\partial {y_l}}}} \right)}^2}} \right)^2} + {\left( {\frac{{{\Psi _i}{\Theta _i}}}{{{r_i}}}} \right)^2} + {\left( {\frac{{2\pi }}{\lambda }\frac{{\partial {r_{i,l}}}}{{\partial {y_l}}}\frac{{\partial {r_{i,l}}}}{{\partial {z_l}}}} \right)^2} + \\
 &\left. {{{\left( {\frac{{{\Theta _i}{\Phi _i}}}{{{r_i}}}} \right)}^2} + {{\left( {\frac{{2\pi }}{\lambda }\frac{{\partial {r_{i,l}}}}{{\partial {z_l}}}\frac{{\partial {r_{i,l}}}}{{\partial {x_l}}}} \right)}^2} + {{\left( {\frac{{{\Theta _i}{\Psi _i}}}{{{r_i}}}} \right)}^2} + {{\left( {\frac{{2\pi }}{\lambda }\frac{{\partial {r_{i,l}}}}{{\partial {z_l}}}\frac{{\partial {r_{i,l}}}}{{\partial {y_l}}}} \right)}^2} + {{\left( {\frac{{1 - \Theta _i^2}}{{{r_i}}}} \right)}^2} + {{\left( {\frac{{2\pi }}{\lambda }{{\left( {\frac{{\partial {r_{i,l}}}}{{\partial {z_l}}}} \right)}^2}} \right)}^2}} \right]\\
 &={\left( {\frac{{2\pi }}{\lambda }\sum\limits_{l' = 1,l' \ne l}^L {2{{\left| {{\alpha _i}} \right|}^2}\left| {{V_{k,l,l'}}} \right|} } \right)^2}\left[ {\frac{2}{{r_i^2}} + {{\left( {\frac{{2\pi }}{\lambda }} \right)}^2}{{\left( {1 + \frac{{x_l^2 + y_l^2 + z_l^2 - {{\left( {{x_l}{\Phi _i} + {y_l}{\Psi _i} + {z_l}{\Theta _i}} \right)}^2}}}{{r_i^2}}} \right)}^2}} \right].
 \end{aligned}
 \end{equation}
 \hrulefill
 \end{figure*}
 Let ${C_{\max }} \triangleq \mathop {\max }\limits_{{{\bf{q}}_l}} x_l^2 + y_l^2 + z_l^2$, it follows that  
 \begin{equation}
 \begin{aligned}
 \left\| {{\nabla ^2}{g_{k,i,l}}} \right\|_F^2& \le {\left( {\frac{{2\pi }}{\lambda }\sum\limits_{l' = 1,l' \ne l}^L {2{{\left| {{\alpha _i}} \right|}^2}\left| {{V_{k,l,l'}}} \right|} } \right)^2} \times \\
 &\ \ \ \ \ \left[ {\frac{2}{{r_i^2}} + {{\left( {\frac{{2\pi }}{\lambda }} \right)}^2}{{\left( {1 + \frac{{{C_{\max }}}}{{r_i^2}}} \right)}^2}} \right].
 \end{aligned}
 \end{equation}

 With ${\nabla ^2}{g_{k,i,l}}  \preceq  {\left\| {{\nabla ^2}{g_{k,i,l}}} \right\|_2}{\bf{I}}$, and by choosing ${\delta _{k,i,l}} = \frac{{2\pi }}{\lambda }\sum\limits_{l' = 1,l' \ne l}^L {2{{\left| {{\alpha _i}} \right|}^2}\left| {{V_{k,l,l'}}} \right|} \sqrt {\frac{2}{{r_i^2}} + {{\left( {\frac{{2\pi }}{\lambda }} \right)}^2}{{\left( {1 + \frac{{{C_{\max }}}}{{r_i^2}}} \right)}^2}}$, we have ${\nabla ^2}{g_{k,i,l}}  \preceq  {\delta _{k,i,l}}{\bf{I}}$. Thus, the lower and upper bounds of ${{g_{k,i,l}}}$ in \eqref{lowerBoundgkim} and \eqref{upperBoundgkim} can be obtained with the Taylor's theorem \cite{sun2016majorization}. The proof of Lemma \ref{quadraticSurrogatelemma} is thus completed.

\section{Proof of Proposition \ref{Convergence}}\label{proofOfSingleAntennaUAVConvergence}
 Denote by $\{ {{\bf{q}}_l^{\left( j \right)}}\}$ and $\{ {{\bf{v}}_k^{\left( j \right)}}\}$ the corresponding optimization variables in the $j$-th iteration. Let $\rho _{{\rm{lb}}}^{{\rm{traj}}}( {\{ {{\bf{q}}_l^{\left( j \right)}}\},\{ {{\bf{v}}_k^{\left( j \right)}} \}})$ denote the corresponding objective value of problem \eqref{subProblemTrajectory3}. In the $j$-th iteration, since the optimal solution to \eqref{subProblemBeamforming} is obtained for given $\{ {\bf{q}}_l^{\left( j \right)}\}$ in step 3, it follows that 
 \begin{equation}\label{receiveBeamformingInequality}
 \rho \left( {\left\{ {{\bf{q}}_l^{\left( j \right)}} \right\},\left\{ {{\bf{v}}_k^{\left( j \right)}} \right\}} \right) \le \rho \left( {\left\{ {{\bf{q}}_l^{\left( j \right)}} \right\},\left\{ {{\bf{v}}_k^{\left( {j + 1} \right)}} \right\}} \right).
 \end{equation}
 
 Besides, regarding UAV $l$ in step 5, for given $\{ {\bf{v}}_k^{\left( {j + 1} \right)}\} $ and $\{ {\bf{q}}_1^{(j + 1)}, \cdots ,{\bf{q}}_{l - 1}^{(j + 1)},{\bf{q}}_l^{(j)}, \cdots ,{\bf{q}}_L^{(j)}\}$, we have 
 \begin{equation}\label{placementInequality}
 \begin{aligned}
 &\rho \left( {\left\{ {{\bf{q}}_1^{(j + 1)}, \cdots ,{\bf{q}}_{l - 1}^{(j + 1)},{\bf{q}}_l^{(j)}, \cdots ,{\bf{q}}_L^{(j)}} \right\},\left\{ {{\bf{v}}_k^{\left( {j + 1} \right)}} \right\}} \right)\\
 &\mathop  = \limits^{\left( a \right)} \rho _{{\rm{lb}}}^{{\rm{traj}}}\left( {\left\{ {{\bf{q}}_1^{(j + 1)}, \cdots ,{\bf{q}}_{l - 1}^{(j + 1)},{\bf{q}}_l^{(j)}, \cdots ,{\bf{q}}_L^{(j)}} \right\},\left\{ {{\bf{v}}_k^{\left( {j + 1} \right)}} \right\}} \right)\\
 &\mathop  \le \limits^{\left( b \right)} \rho _{{\rm{lb}}}^{{\rm{traj}}}\left( {\left\{ {{\bf{q}}_1^{(j + 1)}, \cdots ,{\bf{q}}_{l - 1}^{(j + 1)},{\bf{q}}_l^{(j + 1)}, \cdots ,{\bf{q}}_L^{(j)}} \right\},\left\{ {{\bf{v}}_k^{\left( {j + 1} \right)}} \right\}} \right)\\
 &\mathop  \le \limits^{\left( c \right)} \rho \left( {\left\{ {{\bf{q}}_1^{(j + 1)}, \cdots ,{\bf{q}}_{l - 1}^{(j + 1)},{\bf{q}}_l^{(j + 1)}, \cdots ,{\bf{q}}_L^{(j)}} \right\},\left\{ {{\bf{v}}_k^{\left( {j + 1} \right)}} \right\}} \right),
 \end{aligned}
 \end{equation}
 where $\left(a \right)$ holds because the Taylor expansions are tight at the given local point ${\bf{q}}_l^{(j)}$; the inequality $\left( b \right)$ holds because problem \eqref{subProblemTrajectory3} is solved via SCA, which ensures a non-decreasing objective value; the inequality $\left( c \right)$ holds because the objective value of \eqref{subProblemTrajectory1} is lower-bounded by that of \eqref{subProblemTrajectory3} at ${\bf{q}}_l^{(j + 1)}$. Based on \eqref{receiveBeamformingInequality} and \eqref{placementInequality}, we have
 \begin{equation}
 \rho \left( {\left\{ {{\bf{q}}_l^{\left( j \right)}} \right\},\left\{ {{\bf{v}}_k^{\left( j \right)}} \right\}} \right) \le \rho \left( {\left\{ {{\bf{q}}_l^{\left( {j + 1} \right)}} \right\},\left\{ {{\bf{v}}_k^{\left( {j + 1} \right)}} \right\}} \right).
 \end{equation}
 The result shows that the objective value of problem \eqref{OptimizationProblemSingleAntenna} is non-decreasing over iterations, and thus, Algorithm~\ref{alg1} is guaranteed to converge. This thus completes the proof of Proposition \ref{Convergence}. 
  
\end{appendices}


\bibliographystyle{IEEEtran}
\bibliography{refUAVMA}

\end{document}